\documentclass[12pt]{article} 

\usepackage{a4wide}
\usepackage{graphicx} 
\usepackage{amsmath}
\usepackage{amsfonts}
\usepackage{amssymb} 

\newcommand{\nn}{\nonumber}
\newcommand{\be}{\begin{equation}}
\newcommand{\ee}{\end{equation}}
\newcommand{\ba}{\begin{eqnarray}}
\newcommand{\ea}{\end{eqnarray}}
\newcommand{\ci}[1]{\cite{#1}}
\def\vk{{\bf k}_{\perp}}

\def\vb0{{\bf b}_0}

\def\gev{\,{\rm GeV}}

\def\xbj{x_{\rm B}}

\newcommand{\wf}{wave function}
\newcommand{\lsim}{\raisebox{-4pt}{$\,\stackrel{\textstyle
                                                         <}{\sim}\,$}}
\newcommand{\gsim}{\raisebox{-4pt}{$\,\stackrel{\textstyle
                                                         >}{\sim}\,$}}

\newcommand{\tw}{\textwidth}
                                                         
\newcommand{\req}[1]{(\ref{#1})}

\def\={\,=\,}

\def\veps{\varepsilon}

\begin{document}

\thispagestyle{empty}
\begin{flushright}
IRFU-12-174 \\
WUB/12-22 \\
January, 2 2013\\[20mm]
\end{flushright}

\begin{center}
{\Large\bf From hard exclusive meson electroproduction to deeply virtual
  Compton scattering} \\
\vskip 10mm

{\large P.~Kroll} \footnote{Email:  kroll@physik.uni-wuppertal.de}
\\[1em]
{\small {\it Fachbereich Physik, Universit\"at Wuppertal, D-42097 Wuppertal,
Germany}}\\
and\\
{\small {\it Institut f\"ur Theoretische Physik, Universit\"at
    Regensburg, \\D-93040 Regensburg, Germany}}\\
\vskip 10mm

{\large H.~Moutarde} \footnote{Email:  herve.moutarde@cea.fr}, {\large F.~Sabati\'e} \footnote{Email:  franck.sabatie@cea.fr}
\\[1em]
{\small {\it IRFU/Service de Physique Nucl\'eaire, CEA Saclay, F-91191 Gif-sur-Yvette, France}}\\

\end{center}
\vskip 5mm 
\begin{abstract}
  We systematically evaluate observables for hard exclusive 
electroproduction of real photons and compare them to experiment using 
a set of Generalized Parton Distributions (GPDs) whose parameters are constrained by 
Deeply Virtual Meson Production data, nucleon form factors and parton distributions. 
The Deeply Virtual Compton Scattering amplitudes are calculated 
to leading-twist accuracy and leading order in QCD perturbation theory 
while the leptonic tensor is treated exactly, without any 
approximation.
This study constitutes a check of the universality of the GPDs. We summarize 
all relevant details on the parametrizations of the GPDs and describe
its use in the handbag approach of the aforementioned hard scattering
processes. We observe a good agreement between predictions and measurements 
of deeply virtual Compton scattering on a wide kinematic range, including most data 
from H1, ZEUS, HERMES, Hall~A and CLAS collaborations for unpolarized and polarized 
targets when available. We also give predictions 
relevant for future experiments at COMPASS and JLab after the 12~GeV upgrade.
\end{abstract}

\vskip 10mm

\section{Introduction}
\label{intro}
For the last 15 years, the handbag approach to hard exclusive leptoproduction
of photons (DVCS)\footnote{Formally, Deeply Virtual Compton Scattering or DVCS refers only
to the sub-process $\gamma^* p \to \gamma p$. However, DVCS is often used more loosely
in the literature to name the photon leptoproduction process $l p \to l p \gamma$ used experimentally.}
and mesons (DVMP) off protons has been extensively 
investigated both theoretically and experimentally. The handbag approach 
bases on factorization into hard (short-distance) partonic subprocesses 
and soft (long-distance) hadronic matrix elements \ci{rad96,ji96,col96,
collins99}. The latter are parametrized in terms of Generalized Parton 
Distributions (GPDs) \ci{ji96,muller94,rad97}. The GPDs encode information 
on the longitudinal momentum distributions of the partons inside the proton 
as well as on the transverse localization of the partons~\cite{Burkardt:2000za,Diehl:2002he,Ralston:2001xs}.
The forward limits 
of some of the GPDs are the usual parton distributions (PDFs) and the lowest 
moments of the quark GPDs are related to the form factors of the proton. The 
GPDs give access to the total angular momenta of the partons making up 
the proton via Ji's sum rule \ci{ji96}. Another important property of the 
GPDs is their universality, \textit{i.e.} the same GPDs occur in DVCS 
as well as in DVMP although in different flavor combinations.   
At present analytic methods to compute GPDs from QCD are lacking. Only lattice 
QCD provides numerical results on the lowest few moments of $u$ and $d$ quark
GPDs \ci{Bratt:2010jn,hagler11} for unphysical pion masses. There are also a 
number of models for GPDs available (\textit{e.g.} Refs.~\ci{goeke,DFJK1,scopetta,pasquini})
which can be confronted to experimental
data on hard exclusive reactions in order to learn about GPDs.

In the late nineties when the phenomenology of hard exclusive reactions
commenced, estimates of observables for such reactions were made on the 
basis of simple ansaetze for the GPDs, see for instance Refs.~\ci{man97}--\ci{freund01}.  
As it turned out in the course of time, these GPD models were insufficient to
account for the increasing amount of accurate data coming from HERMES and Jefferson Lab.
More complex parametrizations of the GPDs were invented and utilized to 
analyze independently the data on DVCS \ci{kume07,kume09} as well as DVMP 
for light vector mesons \ci{GK1,GK3}  and for pions \ci{liuti09,GK5}. It 
should be noted that in DVCS frequently only reduced amplitudes, the 
so-called Compton Form Factors (CFF), have been extracted \ci{guidal09,moutarde09}
which represent valuable constraints on the GPDs.

It is only recently that the universality property of GPDs has fully been
exploited and a combined analysis of DVCS and DVMP carried
through or a set of GPDs extracted from either DVCS or DVMP used to evaluate
the other reaction. Thus, for instance, Me\v{s}kaukas and M\"uller \ci{mes11}
performed a combined analysis of the HERA data on DVCS and DVMP.
These authors also
used the GPD $H$ advocated for in Refs.~\ci{GK1,GK3} from an analysis of DVMP
data, to compute DVCS observables for HERA kinematics. For a first attempt
to compute DVCS along the same lines also for other kinematical regions
see M\"uller {\it et al} in Ref.~\ci{EIC}. The results look quite promising, no
severe discrepancy has been observed by these authors. In the present article
we are going to study systematically predictions for DVCS in an 
leading-order (LO), lowest-twist calculation of the DVCS amplitudes
using GPDs determined in Refs.~\ci{GK1,GK3,GK5}. In contrast to earlier work, 
\textit{e.g.} Ref.~\ci{belitsky01},  
 the leptonic tensor is evaluated without any 
approximation as it is done in Ref.~\cite{Belitsky:2010jw} implying the 
inclusion of effects that are suppressed by powers of $1/Q$ in our analysis.
The primary goal of our study is to examine how realistic
the considered set of GPDs is in order to eventually improve it if necessary.

The plan of the paper is as follows: in Sect.~\ref{sec:2} we recapitulate the
parametrization of the GPDs under scrutiny, and the theoretical description of
the DVMP and DVCS processes in the handbag approach. In the second part,
Sect.~3, we systematically compare model expectations to existing DVCS
measurements. In Sect.~4 we give some predictions for observables which will
be measured by the COMPASS, CLAS and Hall~A collaborations in the near future. 
Finally, our summary and an outlook are presented in Sect.~5.
 
\section{Theoretical description}
\label{sec:2}

\subsection{The parametrization of the GPDs}
\label{subsec:2.1}
In this section we recapitulate the parametrization of the GPDs used in
Refs.~\ci{GK1,GK3,GK5} and specify all the ingredients required to fit the DVMP 
data. The GPDs are functions of three variables, the usual invariant 
momentum transfer, $t$, the skewness defined as the ratio of light-cone
plus components of the incoming ($p$) and outgoing ($p^\prime$) proton 
momenta
\be
\xi\= \frac{(p-p^\prime)^+}{(p+p^\prime)^+} \,.
\ee 
In the generalized Bjorken regime of large $Q^2$, large $W$ but fixed $\xbj$,
it is related to Bjorken-$x$, $\xbj=Q^2/(2p\cdot q)$ by :
\be
\xi \simeq \frac{\xbj}{2-\xbj}\, ,
\label{eq:xi}
\ee 
\noindent where $q$ is the momentum of the virtual photon and $Q^2$ its virtuality.
The generalized Bjorken regime is defined by large $Q^2$ and large photon-proton cms energy
$W$, but fixed $x_B$.
The third variable, $x$, represents the average momentum fraction the emitted
($k$) and reabsorbed ($k^\prime$) partons carry with respect to the
average proton momenta
\be
x \= \frac{(k+k^\prime)^+}{(p+p^\prime)^+}\,.
\ee
GPDs further depend on a factorization scale \ci{muller94}, which is usually
taken as the photon virtuality unless specified otherwise. At LO, leading-twist accuracy,
only the GPDs $F=H, E, \widetilde{H}, \widetilde{E}$ contribute to DVCS which
are characterized by the fact that the emitted and reabsorbed partons possess 
the same helicity. In Refs.~\ci{GK1,GK3,GK5} an integral representation of the GPDs 
is used
\be
F^i(x,\xi,t)\=\int_{-1}^{1}\, d\rho\,\int_{-1+|\rho|}^{1-|\rho|}\, d\eta
              \,\delta(\rho+\xi\eta -x)\, f_i(\rho,\eta,t)
              + D_i(x,t)\, \Theta(\xi^2-x^2)\,,
\label{eq:int-rep}
\ee  
where $f_i$ is a double distribution \ci{muller94,rad98}\footnote{The variables $\rho$ 
and $\eta$ are usually denoted by $\beta$ and $\alpha$, respectively. However, we  do 
not use this notation here in order to avoid a clash of notation. These latter symbols 
are already used for powers in the functional form.}
and $D_i$ is the so-called $D$-term \ci{pol99} which appears for the gluon and flavor-singlet
quark combination of the GPDs $H$ and $E$. The label $i$ refers to specific
quark flavors (or appropriate combinations) or to gluons. The $D$-terms only 
contribute to the real parts of the amplitudes. At small skewness however,
the amplitude is dominated by the imaginary part. In Refs.~\ci{GK1,GK3,GK5} the
$D$-terms are neglected. The advantage 
of the double-distribution ansatz for the GPDs is that polynomiality of the 
GPDs is automatically satisfied. For the GPDs $H$ and $E$ the $D$-term ensures 
the appearance of the highest power of the skewness in the Mellin moments 
of the GPDs. 

It is popular to write a double distribution as a product of a
zero-skewness GPD and a weight function \ci{rad98} 
\be
w_i(\rho,\eta)\= \, \frac{\Gamma (2n_i+2)}{2^{2n_i+1}\Gamma^2 (n_i+1)} \,
\frac{[(1-|\rho |)^2-\eta^2]^{n_i}}{(1-|\rho |)^{2n_i+1}} \, ,
\label{eq:weight}
\ee   
that generates the $\xi$ dependence of the GPD~:
\be
f_i(\rho,\eta,t) \= F^i(\rho,\xi=0,t)\, w_i(\rho,\eta)\,.
\label{eq:DD}
\ee
In Refs.~\ci{GK1,GK3,GK5} the parameter $n_i$ is taken as 1 for valence quarks and 
as 2 for sea quarks and gluons. The zero-skewness GPD is parametrized as the 
forward limit of that GPDs multiplied by an exponential in $t$ 
\be
F^i(\rho,\xi=0,t) \= F^i(\rho,\xi=0,t=0)\, \exp \big( t p_{fi}(\rho) \big)
\label{eq:zero-skewness}
\ee
The profile function, $p_{fi}(\rho)$, is parametrized in a Regge-like manner
\be
p_{fi}(\rho) \= -\alpha_{fi}^\prime \ln{\rho} + b_{fi}
\label{eq:profile}
\ee
where $\alpha^\prime$ represents the slope of an appropriate Regge trajectory
and $b$ parametrizes the $t$ dependence of its residue. In Ref.~\ci{DFJK4} a more
complicated profile function for valence quarks has been proposed
\be
p_{fi}(\rho) \= \big(\alpha_{fi}^\prime \ln{1/\rho} + b_{fi}\big)\,(1-\rho)^3 
             + A_{fi}\,\rho(1-\rho)^2
\label{eq:profile-dfjk4}
\ee
and exploited in an analysis of nucleon form factors. At small $x$ the
Regge-like part dominates while, for $x\to 1$, the term $\propto A_{fi}$ takes
the lead. It turned out that there is a correlation between $x$ and $t$: 
The small (large) $x$ behavior of the profile function controls the nucleon form factors at 
small (large) $-t$. For parametrizations like \req{eq:zero-skewness} without nodes except at the end-points,
this correlation also holds for other moments of the GPDs and even for convolutions with hard scattering amplitudes.
Thus, the Regge-like profile function is a sufficiently
accurate approximation at small $-t$, the region we are interested in. It 
has also been shown in Refs.~\ci{Burkardt:2002hr, DFJK4} that, at large $x$ the Regge-like profile 
function leads to an unphysically large distance between the struck parton 
and the cluster of the spectators. This distance provides an estimate of the
size of the proton as a whole. 

The decomposition of the quark double distribution $f_q$ for the flavors
$q=u$ and $d$ into $f_{\rm val}^q$ and $f_{\rm sea}^q$ is done following the
convention defined in Ref.~\cite{DiehlReport} :
\begin{eqnarray}
f^q_{\rm val}(\rho,\eta,t) & = & \big[ f_q(\rho,\eta,t) + 
       \epsilon_f f_q(\rho,\eta,t) \big]\, \Theta (\rho ) \nn   \, , \\
f^q_{\rm sea}(\rho,\eta,t) & = & f_q(\rho,\eta,t)\, \Theta (-\rho ) -
\epsilon_f f_q(-\rho,\eta,t)\, \Theta (\rho ) \, .
\label{eq:val-sea}
\end{eqnarray}
where $\epsilon_f=+1$ for $F=H$ and $E$ and $-1$ for $\widetilde{H}$ and
$\widetilde{E}$. For $H$ and $\widetilde{H}$ and $t=0$ this prescription 
corresponds to the usual decomposition of parton distributions into valence
quark and sea quark distributions, e.g. $q_{\rm val}=q-\bar{q}$. The GPDs
respect the symmetry relations
\ba
F^g(-x,\xi,t) &=& \epsilon_f\, F^g(x,\xi,t)\,, \nn\\
F^q_{\rm sea}(-x,\xi,t) &=& -\epsilon_f\, F^q_{\rm sea}(x,\xi,t)\,,
\ea
and 
\be
F^g_{\rm val}(-x,\xi,t) \= 0\,, \qquad -1\leq x\leq -\xi\,.
\ee
For DVCS the C-parity even combination of the double distributions is
required
\be
f^{q (+)}(\rho,\eta,t) \= f_q(\rho,\eta,t) -\epsilon_f\, f_q(-\rho,\eta,t)\,.
\ee

The parameters appearing in the double distributions, in particular in
Eq.~\req{eq:zero-skewness} and Eq.~\req{eq:profile}, are fixed in an analysis of
DVMP data in the kinematical region specified by $\xi\lsim 0.1$, $Q^2\gsim
3\,\gev^2$, $W\gsim 4\,\gev$ and $-t \lsim 0.6\,\gev^2$. 
Cross section data are the only measurements available on a large $Q^2$ range
(up to about $100\,\gev^2$) and they are dominated by contributions from the
GPD $H$. Other GPDs significantly enter asymmetries which are only
measured for $2 \lsim Q^2 \lsim 4\,\gev^2$. Therefore evolution effects in this study are 
only sizeable for the GPD $H$ and can safely be neglected for other 
GPDs \ci{GK3,GK5}. Since the scale dependence of $\widetilde H$ is available in~\cite{GK3},
we still make use of it in our analysis of DVCS although it has practically no bearing on our results.
It should be mentioned that in~\cite{GK1,GK3} the evolution of the GPDs $H$ and $\widetilde H$ is treated
in an approximate way through the evolution of the PDFs in \req{eq:zero-skewness}. A possible evolution
of the profile function is ignored. At least for small skewness and small $-t$ this approximation is
reasonable as has been demonstrated in \cite{GK3}. We follow this recipe which in any case is only of
importance for the description of the DVCS cross section at HERA energies.
It has been checked in Refs.~\ci{GK3,GK5} that the valence
quark GPDs are in agreement with the nucleon form factors at small $-t$
and that all GPDs respect various positivity bounds \ci{DFJK4,pob02,burkardt03}. 
At small $-t$ there is also reasonable agreement between the moments of these
GPDs and recent lattice results \ci{Bratt:2010jn,hagler11}. But the GPD
moments from lattice QCD have a flatter $t$ dependence than those obtained
from the GPDs we are discussing here and also flatter than nucleon form factor
data exhibit. A possible explanation comes from the unphysical values 
of the pion mass used in present lattice simulations and the contamination
from excited states \ci{Green:2012ud}. Chiral extrapolations of the lattice 
moments have not yet been systematically performed, see for instance Ref.~\ci{bali2012}.

\subsubsection{Parametrization of the GPD $H$}
\label{subsec:2.1.1}
The GPD $H$ is rather well determined since it controls the cross sections
for electroproduction of vector mesons for which a wealth of data is available.
An advantage is that its forward limit occurring in Eq.~\req{eq:zero-skewness}, is
a usual PDF. Therefore, only the parameters appearing in the profile functions 
have to be fixed. Linear Regge trajectories are assumed
\begin{equation}
\alpha_{hi} = \alpha_{hi}(0) + \alpha_{hi}' t  \;\;\;\;\;\; {\rm with} \; 
i= {\rm g, sea, val}  
\end{equation}
As it is well-known the intercept controls the low-$x$ behavior of the PDF
\ci{landshoff71}. A standard Regge trajectory is assumed for the valence
quarks, see Tab.\ \ref{table:H}. Since the sea-quark PDF is mainly driven
by evolution for $Q^2\gsim 4\,\gev^2$ it is furthermore assumed that 
$\alpha_{h {\rm sea}}(t)\equiv\alpha_{hg}(t)$. The gluon trajectory with an
effective scale-dependent intercept $\alpha_{hg}(0,Q^2)$, is directly
seen in the HERA experiments \ci{H1-09,zeus07} (and references therein) and
consequently fixed by these data. This trajectory is also quoted in Tab.\ 
\ref{table:H}. The trajectories are accompanied by Regge residues assumed to 
have an exponential $t$ dependence, see Eq.~\req{eq:profile}, with slopes taken as
($m$ being the proton mass)
\begin{eqnarray}
b_{h{\rm val}}&=&0 \, , \nonumber \\
b_{hg}=b_{h{\rm sea}} &=&2.58 \,\gev^{-2} + 0.25 \,\gev^{-2} 
          \ln{\frac{m^2}{Q^2+m^2}}  \,.
\label{eq:slopes-H}
\end{eqnarray}

It is convenient to expand the forward limits of $H$, the PDFs, in a power
series of $\sqrt{x}$ (for $\rho>0$):
\ba
H^g(\rho,\xi=t=0) &=& \rho^{-\delta_g}\, (1-\rho)^5\,
          \sum_{j=0}^3 c_{gj}\, \rho^{j/2}\,, \nn\\
H^i(\rho,\xi=t=0) &=& \rho^{-\alpha_{hi}(0)} (1-\rho )^{2n_i+1} \sum_{j=0}^3
  c_{ij}\, \rho^{j/2} \,, 
\label{eq:PDFexp}
\ea
where, with regard to the fact that the forward limit of $H^g$ is 
defined as $\rho g(\rho)$, 
\be
\delta_g\=\alpha_{hg}(0)-1\,.
\label{eq:intercept}
\ee
The expansion coefficients $c_{ij}=c_{ij}(Q^2)$ have been obtained from a fit 
to the CTEQ6M PDFs \cite{cteq6m}; they are compiled in Tab.\ \ref{table:H} too.

The advantage of this expansion is twofold. First the integral
\req{eq:int-rep} can be worked out term by term analytically leading to
a corresponding expansion of the GPDs
\be
H_i(x,\xi,t)\=e^{b_{hi}t} \sum_{j=0}^3 c_{ij} H_{ij}(x,\xi,t)\,.
\ee
The integrals $H_{ij}$ are given explicitly in Ref.~\ci{GK3}. 
Second, the term $\rho^{-\delta_g}$ in Eq.~\req{eq:PDFexp} guarantees that
the longitudinal cross section on $\rho^0$ and $\phi$ electroproduction 
which is given by
\be
\sigma_L\propto W^{4\delta_g(Q^2)}
\ee
at fixed $Q^2$ and small $\xbj$ ($\xi$), is in reasonable agreement with
the HERA data \ci{H1-09,zeus07}. The data used in current PDF analyses,
for instance Refs.~\ci{MRST04,alekhin} or more recent ones \ci{CT10,MRST10,NNPDF}, 
do not constrain the gluon PDF well for $\rho\lsim 0.01$. However,
forcing the expansion of the gluon and sea PDFs to behave as $\rho^{-\delta_g}$ at
low $\rho$ always leads to reasonable agreement of the DVMP cross section 
with the HERA experiments. With this prescription other sets of PDFs,
{\sl e.g.}\ Refs.~\ci{MRST04,alekhin}, provide similar results as CTEQ6M. We stress that in
all cases the expansions \req{eq:PDFexp} are in good agreement with the
original PDFs within their quoted errors.

Finally, in accord with CTEQ6 analysis, the quark sea is simplified in
Ref.~\ci{GK3} as
\begin{eqnarray}
H^u_{\rm sea} &=& H^d_{\rm sea} =  \kappa_s H^s_{\rm sea} \, , \nonumber \\
{\rm with} \;\;\;\;\;\; \kappa_s&=&1+0.68/(1+0.52 \ln Q^2/Q_0^2 ) \, ,
\end{eqnarray}
where the $Q^2$ dependence of the flavor symmetry breaking factor $\kappa_s$ 
was taken from the CTEQ6M PDFs. 

\begin{table*}
\renewcommand{\arraystretch}{1.4} 
\begin{center}
\begin{tabular}{|c||c|c|c|c|}
\hline
& gluon & strange & $u_{\rm val}$ & $d_{\rm val}$ \\ \hline \hline
$\alpha(0)$ & $1.10+0.06\,L-0.0027L^2$ & $\alpha_{hg}(0)$ & 0.48 & 0.48 \\ \hline
$\alpha'$ & $0.15$~GeV$^{-2}$ & 0.15~GeV$^{-2}$ & 0.9~GeV$^{-2}$ & 0.9~GeV$^{-2}$ \\ \hline
$c_0$ & $\phantom{-}2.23+0.362\,L$ & $\phantom{-}0.123+0.0003\,L$ & $1.52+0.248\,L$ & $\phantom{-}0.76+0.248\,L$ \\ 
$c_1$ & $\phantom{-}5.43-7.00\,L$ & $-0.327-0.004\,L$ & $2.88-0.940\,L$ & $\phantom{-}3.11-1.36\,L$ \\ 
$c_2$ & $-34.0+22.5\,L$ & $\phantom{-}0.692-0.068\,L$ & $-0.095\,L$ & $-3.99+1.15\,L$ \\ 
$c_3$ & $\phantom{-}40.6-21.6\,L$ & $-0.486+0.038\,L$ & 0 & 0 \\
\hline
\end{tabular}
\caption{\small Parameters used for the GPD $H$, with $L=\ln{(Q^2/Q_0^2)}$ and 
$Q_0^2=4$~GeV$^2$ for the CTEQ6M PDF set.}
\label{table:H}
\end{center} 
\renewcommand{\arraystretch}{1.0}
\end{table*}

\subsubsection{Parametrization of $E$}
\label{subsec:2.1.2}

Much less is known about $E$ than for $H$. There is only the analysis of
the Pauli form factor \ci{DFJK4} which provides information on $E$ for valence 
quarks. In addition there is an admittedly weak constraint from the
asymmetries in electroproduction of $\rho^0$ mesons measured with a
transversely polarized target \ci{Airapetian:2009ad,Adolph:2012ht}, for
details see Sect.\ \ref{subsec:2.2}.

The GPD $E$ does not reduce to a PDF, the forward limit is not accessible in
DIS. Therefore, the forward limit is to be fixed from exclusive experimental
data as well. It is parametrized like the usual PDFs:
\begin{equation}
E^q_{\rm val}(\rho,\xi=0,t=0) \= B^{-1}(1-\alpha_{\rm val}, 1+\beta^q_{\rm val})
                       \kappa_q \, \rho^{-\alpha_{\rm val}} (1-\rho)^{\beta^q_{\rm val}} \, ,
\end{equation}
where $B(a,b)$ is Euler's beta function. 
The prefactor ensures the correct normalization of the Pauli form factor at $t=0$.
Indeed the $n=1$ moment 
\be
e_{n0}^{q_v}=\int_0^1 d\rho\,\rho^{n-1}\,E^q_{\rm val}(\rho,\xi=t=0)
\ee
reduces to $\kappa_q$ which is the flavor-$q$ contribution to the nucleon 
anomalous magnetic moment ($\kappa_u=1.67$\, , $\kappa_d=-2.03$).  
The fits to the nucleon Pauli form factors performed in Ref.~\cite{DFJK4} fix 
the parameters specifying $E$ for valence quarks:
\begin{equation}
\beta^u_{\rm val}=4 \, , \;\;\;\; \beta^d_{\rm val}=5.6 \, .
\label{eq:beta-E}
\end{equation}
In the spirit of the Regge model, the trajectory $\alpha_{e{\rm val}}$ is 
taken to be the same as in $H$, see Tab.\ \ref{table:H}. The profile
  function \req{eq:profile} is evaluated with 
$\alpha^\prime_{e{\rm val}}=\alpha^\prime_{h{\rm val}}$ and with 
slope parameters $b_{e{\rm val}}$ taken to be zero.

All this specifies the double distribution \req{eq:DD} for $E^q_{\rm val}$.
The GPDs are then obtained from Eq.~\req{eq:int-rep} where the factor 
$(1-\rho)^{\beta^d_{\rm val}-3}$ is expanded in a power series up to order
8 in order to perform the integration analytically. We stress that in the
form factor analysis \ci{DFJK4} the more complicated profile function
\req{eq:profile-dfjk4} has been used while in the analysis of meson
electroproduction \ci{GK1,GK3,GK5} the small-$\rho$ approximation
\req{eq:profile} is adopted. There is an ongoing
reanalysis of the form factor data \ci{DK12}; preliminary results are close to
those of Eq.~\req{eq:beta-E}. The uncertainties in the determination of $E$ is reduced as
compared to the results provided in Ref.~\ci{DFJK4}.

The determination of $E$ for gluons and sea quarks is still in its infancy.
A rough estimate of these GPDs has been made in Ref.~\ci{GK4} along the lines 
proposed by Diehl and Kugler \ci{kugler}. Again the double distribution
construction is used and the forward limits of the gluonic and strange quark
GPDs are parametrized as
\begin{eqnarray}
E^s(\rho,\xi=t=0) &=& N_s \rho^{-1-\delta_g} (1-\rho)^{\beta_{Es}} \, , \nonumber \\
E^g(\rho,\xi=t=0) &=& N_g \rho^{-\delta_g} (1-\rho)^{\beta_{Eg}} \, .
\label{eq:esea-egluon}
\end{eqnarray}
Of course the same Regge trajectory as for $H$ is used, see Tab.\
\ref{table:H} and Eq.~\req{eq:intercept}. The lack of detailed information forces the
assumption of a flavor-symmetric sea. 
The powers of the large $\rho$ behavior are set to the following values 
(variant 3 of Tab.\ 1 in Ref.\ \ci{GK4}): 
\begin{equation}
\beta^s=7 \, , \;\;\;\; \beta^g=6 \, .
\end{equation}
The integer powers allow to solve the integral \req{eq:int-rep} analytically.
The slope of the residues are taken as :
\begin{equation}
b_{eg}=b_{es} =0.9\, b_{hg}\,,
\end{equation}
see Eq.~\req{eq:slopes-H}. In Ref.\ \ci{GK4} the normalization, $N_s$, of $E^s$ is 
fixed from saturating a positivity bound \ci{DFJK4}
 for a certain range of $\rho$. 
Since the bound is quadratic the sign of $N_s$ is not fixed. Therefore,
we have to consider the two cases $N_s=\pm 0.155$.  

The normalization of $E^g$ is fixed by using a sum rule for
the second moments of $E$ \ci{kugler}
\be
e^g_{20}\= - e_{20}^{u_{\rm val}} -e_{20}^{d_{\rm val}} - 2\sum_{i=\bar{u},\bar{d},\bar{s}}
e_{20}^{i}
\label{eq:sum-rule}
\ee
The valence quark contribution to this sum rule is very small. Hence, the
gluon and sea quark moments almost cancel each other.
In any case, the sum rules allows to fix the normalization $N_g$ for given $N_s$.

\subsubsection{The GPD $\widetilde H$}
\label{subsec:2.1.3}

The forward limit of $\widetilde{H}$ reduces to the polarized PDF for
which in Refs.~\ci{GK3,GK5} the Bl\"umlein-B\"ottcher results \ci{BB} are taken.
To fix the parameters of $\widetilde{H}$ only the HERMES data on the cross
sections and the target asymmetries for $\pi^+$ electroproduction \ci{hermes-pi,hermes2} are at disposal.
Therefore,  $\widetilde{H}$ is determined only for the valence quarks,
$\widetilde{H}_{\rm sea}$ and $\widetilde{H}^g$ are neglected.

Analogously to Eq.~\req{eq:PDFexp} $\widetilde{H}^i_{\rm val}$ is expanded in 
a power series
\be
\widetilde{H}^q_{\rm val}(\rho,\xi=t=0)\= 
         \eta_q\,A_q\rho^{-\alpha_{\tilde{h}q}(0)}\,
              (1-\rho)^3\,\sum_{j=0}^2 \tilde{c}_{qj}\,\rho^j\, ,
\ee
where $q=u,d$ and $A_q$ is a normalization factor. The GPDs
are constrained by the lowest moments at $\xi=t=0$
\be
\eta_q \= \int_0^1 d\rho\,\widetilde{H}^q_{\rm val}(\rho,\xi=t=0)\,
\ee 
which are known from $F$ and $D$ values and $\beta$-decay constants in flavor SU(3)
\be
\eta_u\=0.926\pm 0.014\,, \qquad \eta_d\=-0.341\pm 0.018\,.
\ee
This normalization is guaranteed by the factor
\be
A_q^{-1} \= B(1-\alpha_{\tilde{h}q},4)\left[\tilde{c}_{q0}
      +\tilde{c}_{q1}\,\frac{1-\alpha_{\tilde{h}q}}{5-\alpha_{\tilde{h}q}}
    + \tilde{c}_{q2}\, \frac{(2-\alpha_{\tilde{h}q})(1-\alpha_{\tilde{h}q})}
                   {(6-\alpha_{\tilde{h}q})(5-\alpha_{\tilde{h}q})}\right]\,.
\ee
The expansion coefficients are compiled in Tab.\ \ref{table:Htilde}
as well as the other parameters specifying this GPD.

\begin{table*}
\renewcommand{\arraystretch}{1.4} 
\begin{center}
\begin{tabular}{|c||c|c|}
\hline
    & $u_{\rm val}$ & $d_{\rm val}$ \\ \hline \hline
$\alpha(0)$ & 0.48 & 0.48 \\ \hline
$\alpha'$ & 0.45~GeV$^{-2}$ & 0.45~GeV$^{-2}$ \\ \hline
$b_{\tilde{h}}$ & 0  & 0 \\ \hline
$\tilde{c}_0$ & 0.170+0.03\,L  & -0.320-0.040\,L  \\ 
$\tilde{c}_1$ & 1.340-0.02\,L & -1.427-0.176\,L \\ 
$\tilde{c}_2$ & 0.120-0.40\,L & \phantom{-}0.692-0.068\,L  \\ 
\hline
\end{tabular}
\caption{\small Parameters used for the GPD $\widetilde{H}$. Evolution is parametrized through the variable $L=\ln (Q^2/Q_0^2)$ with
 $Q_0^2=4$~GeV$^2$.}
\label{table:Htilde}
\end{center} 
\renewcommand{\arraystretch}{1.0}
\end{table*}

\subsubsection{The GPD $\widetilde E$}
\label{subsec:2.1.4}

Only exclusive $\pi^+$ electroproduction data constrain this GPD in the
DVMP analysis. Therefore, as for $\widetilde{H}$, it can only  be fixed 
for valence quarks. As is well-known it consists of two parts the pion-pole
contribution and a non-pole one. The pole contribution reads \ci{VGG,pentinen}
\be
\widetilde{E}^u_{\rm pole}\=-\widetilde{E}^d_{\rm pole}\=
          \Theta(|x|\leq \xi)\, \frac{F_P(t)}{4\xi}\,
                     \Phi_\pi\Big(\frac{x+\xi}{2\xi}\Big)\,,
\ee
where $F_P$ is the pseudoscalar from factor of the nucleon. With the
help of PCAC and the Goldberger-Treiman relation the pole contribution
to the pseudoscalar form factor can be written as
\be
F_P(t)\= -m_N f_\pi\,\frac{2\sqrt{2} g_{\pi NN} F_{\pi NN}(t)}{t-m_\pi^2}\,.
\ee
Here $m_N$ ($m_\pi$) is the mass of the nucleon (pion), $g_{\pi NN}$ ($=13.1$)
is the pion-nucleon coupling constant, $f_\pi$ is the pion decay constant and
$\Phi_\pi$ is the pion's distribution amplitude taken as:
\be
\Phi_\pi(\tau)\=6\tau(1-\tau)\,\big[1+a_2C^{3/2}_2(2\tau-1)\big]\,.
\ee
The Gegenbauer coefficient is taken as $a_2=0.22$ at the initial scale
$Q_0^2=4\,\gev^2$. This value for $a_2$ is conform with the sharp rise of
the $\pi\gamma$ transition form factor with $Q^2$ as is observed by the Babar
collaboration \ci{BABAR}. However, this behavior of the transition from factor
is not seen by the BELLE collaboration \ci{BELLE}. The BELLE data are
compatible with a pion distribution amplitude close to the asymptotic form
$6\tau(1-\tau)$. Since $\widetilde{E}$ plays a minor role in DVCS we keep the
value 0.22 for $a_2$; it practically has no bearing on our predictions for DVCS.

Finally, $F_{\pi NN}$ is the form
factor of the pion-nucleon vertex and is parametrized in Ref.~\ci{GK5} as
\be
F_{\pi NN} \=\frac{\Lambda_N^2-m_\pi^2}{\Lambda_N^2-t^\prime}
\ee
with $\Lambda_N=0.44\,\gev$. 

The non-pole contribution for which there is only a weak evidence in the data,
is parameterized in the same spirit as the other GPDs. The forward limit
is parametrized as in Refs.~\ci{GK5,GK6}
\be
\widetilde{E}^q_{\rm val}(\rho,\xi=t=0)\=N_{\tilde{e}}^q 
        \rho^{\alpha_{\tilde{e}}(0)}\,(1-\rho)^5\,.
\ee
Flavor independence of the Regge trajectory and the slope of its residue
is assumed.
The parameters for $\widetilde{E}$ are compiled in Tab.\ \ref{table:Etilde}.
\begin{table*}
\renewcommand{\arraystretch}{1.4} 
\begin{center}
\begin{tabular}{|c|c|c||c|c|}
\hline
    $\alpha_{\tilde{e}}(0)$ & $\alpha^\prime_{\tilde{e}}$ & $b_{\tilde{e}}$
    ($\gev^{-2}$) & $N^u_{\tilde{e}}$ & $N^d_{\tilde{e}}$ \\ \hline 
0.48 & 0.45 & 0.9 & 14.0 & 4.0 \\\hline
\end{tabular}
\caption{\small Parameters used for the GPD $\widetilde{E}$. Evolution
is ignored.}
\label{table:Etilde}
\end{center} 
\renewcommand{\arraystretch}{1.0}
\end{table*}

\subsection{Description of DVMP}
\label{subsec:2.2}

The particular variant of the handbag approach used to extract GPDs from 
 meson electroproduction in the kinematical region of large $Q^2$ and 
large $W$ but small $\xbj (\lsim 0.2)$ and small invariant momentum transfer 
$-t$, is described in some detail in \ci{GK1,GK3}. Here, only the basic facts 
are reviewed. As an example, let us examine the helicity amplitudes for the asymptotically leading
transitions from longitudinally polarized photons to likewise polarized $\rho^0$
mesons, $\gamma^*_L p\to\rho^0_L p$ :
\ba
{\cal M}_{0+, 0 +} &=& \frac{e}{2} \frac1{\sqrt{2}} 
\left\{ {\cal H}^g_{\rm V,eff} 
          +  e_u {\cal H}^u_{\rm V,eff}-e_d {\cal H}_{\rm V,eff}^d\right\}\,,\nn\\
{\cal M}_{0 -, 0 +} &=& -\frac{e}{2}\frac{\sqrt{-t^\prime}}{2m_N}\frac1{\sqrt{2}}
\left\{ {\cal E}_V^g +  e_u {\cal E}_V^u -e_d {\cal E}_V^d \right\}\,,
\label{hel-amp}
\ea
within the handbag approach. The helicities of the protons are labeled by
their signs and $e_q$ denotes the charge of the quark with flavor $q$ in units
of the positron charge $|e|$.  To the proton helicity-non-flip amplitude the
GPDs contribute in the combination
\be
H_{\rm eff} \= H - \frac{\xi^2}{1-\xi^2} E\,.
\ee
The terms ${\cal F}_{\rm V}$ in Eq.~\req{hel-amp} denote convolutions of 
subprocess amplitudes and GPDs $F$ ($=H, E$):
\be
{\cal F}^{\,i}_{\rm V}(\xi, t, Q^2) \= \sum_\lambda \int^1_{x_i} dx 
            {\cal A}^i_{0\lambda, 0\lambda}(x,\xi,Q^2,t=0)\, F^i(x,\xi,t)
\label{eq:convolutions-mesons}
\ee
where $i=g, q$, $x_g=0$ and $x_q=-1$. The subprocess amplitude ${\cal A}$
for partonic helicity $\lambda$ is to be calculated perturbatively.
Since $E$ is of the same order as $H$ in absolute value (see the discussion in
Sec.\ \ref{subsec:2.1}), one can approximate $H_{\rm eff}$ by $H$ in the small 
skewness region. Moreover, for small $-t^\prime$ the helicity-flip amplitude 
can also be neglected in the cross sections for vector mesons ( an exception 
is $\rho^+$ production) which is therefore only sensitive to $H$. 

\begin{figure}[t]
\begin{center}
\includegraphics[width=0.45\tw]{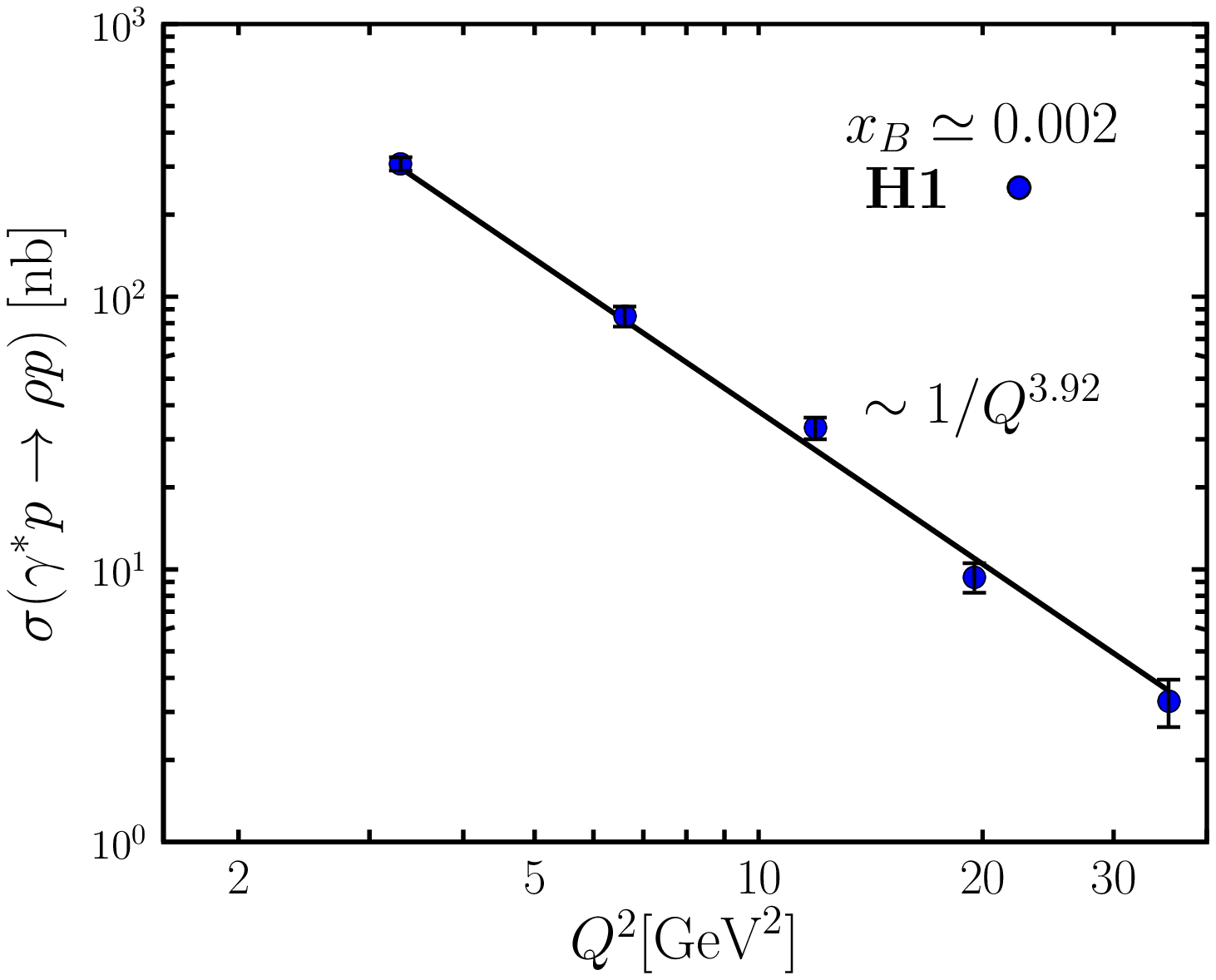} 
\includegraphics[width=.45\tw]{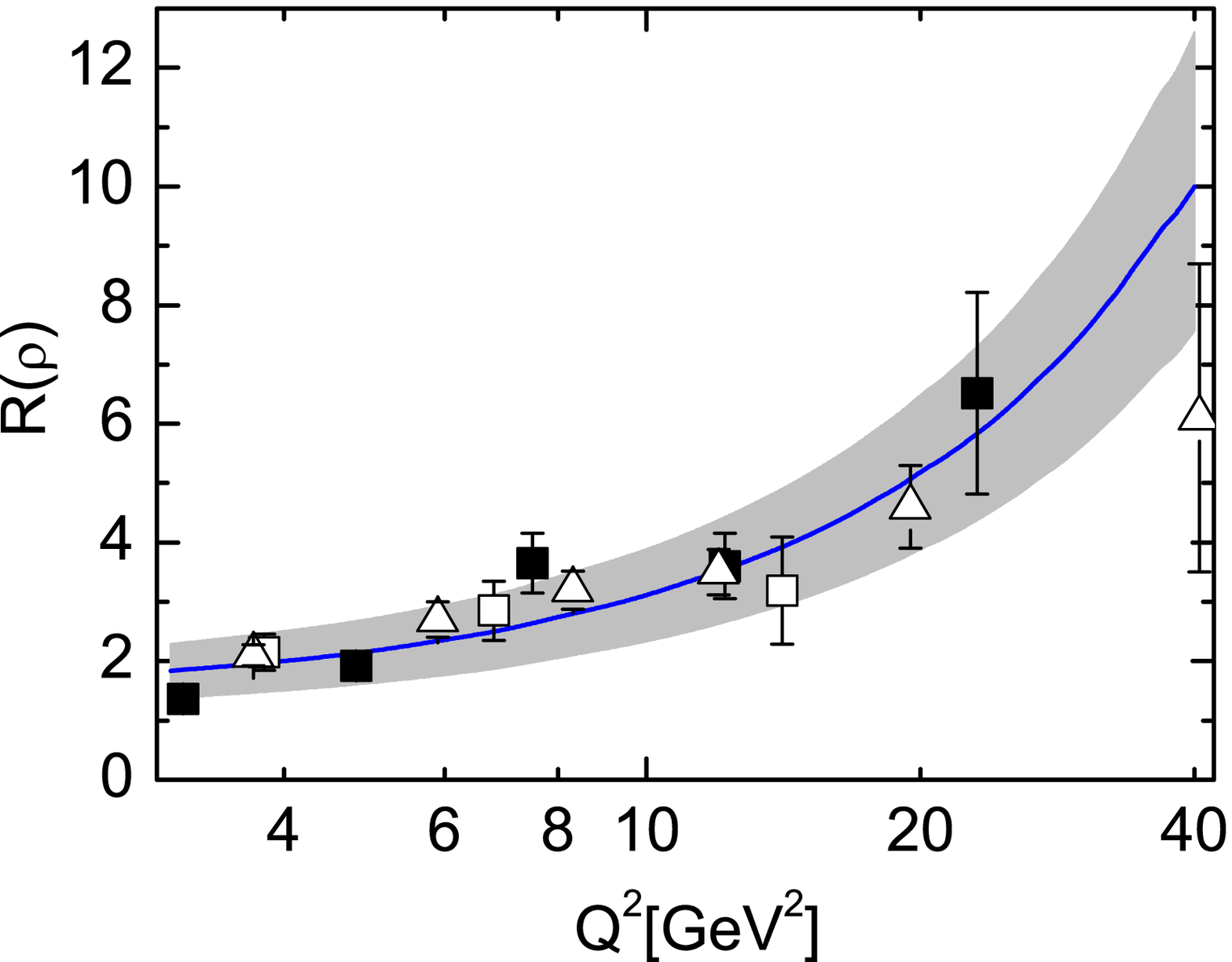}
\caption{Left: The cross section for $\rho^0$ electroproduction vs. $Q^2$ at
  $\xbj\simeq 0.002$. Data are taken from Ref.~\ci{H1-09} and compared to a power-law
  fit. Right: The ratio of $\sigma_L$ and $\sigma_T$ for $\rho^0$ production
  vs. $Q^2$ at $W=90\,\gev$. Data taken from Refs.~\ci{H1-09,zeus07}. The figure 
is taken from Ref.~\ci{GK3} where also further references to data can be found.}
\label{fig:sigma}
\end{center}
\end{figure}

As is well-known, in collinear factorization the handbag result for the 
integrated cross section scales as $1/Q^6$ at fixed $x_B$. However,
for the kinematics accessible to current experiments, the data are in conflict
with this prediction. This can be seen from Fig.\ \ref{fig:sigma} --
the recent H1 data \ci{H1-09} on $\rho^0$ production only drop as 
$\simeq 1/Q^4$. The theoretical $1/Q^6$ behavior of the cross section
is modified by logs of $Q^2$ generated by the evolution of the GPDs; they  
diminish the discrepancy between theory and experiment.
Experiment also tells us that the transverse cross section, 
$\sigma_T$, is not small; the ratio $R=\sigma_L/\sigma_T$, also shown in Fig.\
\ref{fig:sigma}, is rather small for experimentally accessible values of
$Q^2$. Combining the data on the ratio with the unseparated cross section it
becomes clear that the longitudinal cross section approximately falls off
as $\simeq 1/Q^4$ too. This fact implies a marked overestimate of the
longitudinal cross section at $Q^2\simeq 4\,\gev^2$ in the collinear
approximation if it is evaluated from GPDs of the type discussed in Sect.\
\ref{subsec:2.1.1}. Agreement with experiment is however achieved in 
collinear approximation for an alternative parameterization of the GPD $H$ 
proposed in Ref.~\ci{mes11}. The evolution of this GPD produces much larger
logs of $Q^2$ than the GPD described in this article.

In view of this situation,  power corrections are added to the LO subprocess 
amplitudes in Refs.\ \ci{GK1,GK3}. These power corrections are calculated 
within the modified perturbative approach \ci{botts89} 
in which quark transverse degrees of freedom are kept and gluon radiation
is taken into account. The latter has been calculated in the form of a 
Sudakov factor to next-to-leading-log approximation using resummation 
techniques and having recourse to the renormalization group \ci{botts89}. 
For consistency, allowance is to be made for meson light-cone \wf s 
instead of distribution amplitudes. Due to these power corrections the 
convolutions ${\cal F}_V^i$ also depend on $Q^2$. The modified perturbative 
approach is designed in such a way that asymptotically the collinear result 
for the subprocess amplitudes emerges. It is to be stressed that, in contrast
to the situation at the mesonic vertex, the partons entering the subprocess
are treated as being emitted and reabsorbed by the proton collinearly. 

The above described approach can also be applied to vector mesons other than
the $\rho^0$ and can be generalized to the asymptotically suppressed
amplitudes for $\gamma^*_T\to V_T$ transitions\footnote{Another method
to treat $\gamma_T^*\to V_T$ transitions has been
proposed in Ref.~\cite{Anikin:2009bf}}. In collinear approximation
these amplitudes are infrared singular but in the modified perturbative 
approach $\vk$ in the propagators regularizes the singular integrals. 
It should be mentioned that the described approach bears similarities to 
the color dipole model, see Ref.~\ci{nikolaev12} and references therein.
The described approach can also be applied to electroproduction
of pseudoscalar mesons~\cite{GK5,GK6} where one learns about the
valence-quark components of GPDs
$\widetilde {H}$ and $\widetilde{E}$ (see Sections \ref{subsec:2.1.3} and \ref{subsec:2.1.4})
as well as on some of the transversity GPDs.
However, the latter do not contribute to DVCS.

\begin{figure}[t]
\begin{center}
\includegraphics[width=.41\tw]{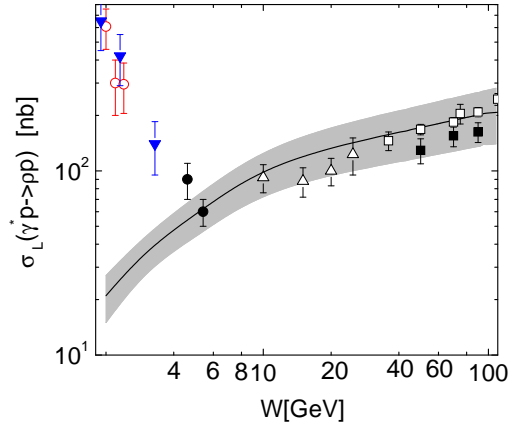}\hspace*{0.05\tw}
\caption{The longitudinal cross section of $\rho^0$ electroproduction vs. $W$ at
  $Q^2=4\,\gev^2$. The handbag result is shown as a solid line; the shadowed 
band represents the uncertainties of this result. For references to the data 
see~\ci{GK3,GK2}. Figure taken from~\cite{GK2}.} 
\label{fig:sigma-rho}
\end{center}
\end{figure}

The available data from HERMES, COMPASS, E665, H1 and ZEUS on cross sections 
and spin density matrix elements for $\rho^0$ and $\phi$ electroproduction
have been analyzed in Refs.\ \ci{GK1,GK3}. The data cover a large range of
kinematics: $Q^2$ varies between 3 and $100\,\gev^2$ and $W$ between 5 and
$180\,\gev$. In Fig.\ \ref{fig:sigma} the ratio $\sigma_L/\sigma_T$ is shown
for $\rho^0$ electroproduction in order to demonstrate that a fair 
description of the amplitude for $\gamma^*_T\to V_T$ is also achieved. 
In Fig.\ \ref{fig:sigma-rho} the longitudinal cross section for $\rho^0$
production is shown versus $W$ at $Q^2=4\,\gev^2$. As the inspection of the 
figure reveals, a good description of all the low-$x_B$ data has been
achieved for $W\gsim 4\,\gev$. For smaller $W$ the handbag results deviate
from experiment; at $W=2\,\gev$ theory and experiment deviate by orders of magnitude.
For $\phi$ production, on the other hand, the handbag seems to work even at 
$W\simeq 2\,\gev$. More results and references to the experimental data can be 
found in Refs.~\ci{GK3,GK2}. The HERMES cross section data on
$\pi^+$ electroproduction data \ci{hermes-pi} as well as various single spin
asymmetries are analyzed in Refs.~\ci{GK5,GK6}.

In the mentioned kinematical region, information on the GPD $E$ can only be
extracted from the target asymmetry 
\be
A_{UT}^{\sin{(\phi-\phi_S)}} \sim {\rm Im} \Big[{\cal E}_{\rm V}^* {\cal H}_{\rm V}\Big]\,,
\label{eq:aut}
\ee 
which is obtained through the $\sin{(\phi-\phi_S)}$ harmonic of the electroproduction 
asymmetry measured with a transversally polarized target. Here, $\phi$ is
the azimuthal angle between the lepton and the hadron planes and $\phi_S$
specifies the orientation of the target spin vector with respect to the lepton plane 
(in the Trento convention~\cite{Bacchetta:2004jz}).  
The convolutions ${\cal E}_{\rm V}$ and ${\cal H}_{\rm V}$ in Eq.~\req{eq:aut} are defined
in Eq.~\req{eq:convolutions-mesons}.  It is shown in Ref.~\ci{GK4} that the GPD $E$
described in Sect.\ \ref{subsec:2.1.2} in combination with the rather
well-known GPD $H$, provides results in agreement with recent $\rho^0$ data on
this asymmetry from the HERMES \ci{Airapetian:2009ad} and COMPASS
\ci{Adolph:2012ht} collaborations. It is to be stressed that for $\rho^0$
production the $\sin{(\phi-\phi_S)}$ harmonic of the cross section for a
transversally polarized target essentially probes $E$ for valence quarks; the
gluon and sea-quark contributions cancel each other to a large extent due to the sum
rule \req{eq:sum-rule} which implies that the second moments of $E^g$ and
$E^{\rm sea}$ have about the same strength but opposite sign. For the simple
parametrization \req{eq:esea-egluon} with no nodes except at the end-points,
this property of the second moments transfers to other moments of $E$ to a
certain degree and in particular to the convolutions. In accord with this
argument, $A_{UT}^{\sin{(\phi-\phi_S)}}$ for electroproduction of the $\phi$
mesons is predicted to be about zero in agreement with preliminary HERMES data
\ci{hermes-aut-phi}.

\subsection{DVCS in the handbag approach}
\label{subsec:handbag}

In the following we use the aforementioned GPDs constrained from DVMP,
nucleon form factors and partons distributions to evaluate $lp\to lp\gamma$
observables and compare to measurements. As for DVMP we will analyze this process 
in the generalized Bjorken regime of large $Q^2$, large $W$ but fixed $\xbj$.

Leptoproduction of photons is complicated since, besides the DVCS contribution,
$\gamma^*p\to \gamma p$, there is also the Bethe-Heitler (BH) contribution. The 
square  of the $l p \rightarrow l p \gamma$ amplitude $\mathcal{M}_{l p \rightarrow l p \gamma}$ 
therefore falls into three parts~:
\begin{equation}
| \mathcal{M}_{l p \rightarrow l p \gamma} |^2 \= | \mathcal{M}_{\textrm{BH}}|^2 
      + \mathcal{M}_{\textrm{I}} + | \mathcal{M}_{\textrm{DVCS}} |^2\,.
\label{eq-bh-interf-vcs}
\end{equation}
These parts readily correspond to the squared amplitudes of the BH and DVCS
processes and their interference.  In the one-photon-exchange approximation of QED, 
the three terms
$|\mathcal{M}_{\textrm{BH}}|^2$, $|\mathcal{M}_{\textrm{DVCS}} |^2$ and $\mathcal{M}_{\textrm{I}}$
in \req{eq-bh-interf-vcs} have the following harmonic structure in $\phi$, the azimuthal 
angle of the outgoing photon with regard to the leptonic plane 
(in the Trento convention~\cite{Bacchetta:2004jz})~:
\begin{eqnarray}
|\mathcal{M}_{\textrm{BH}}|^2 & \propto & \frac{1}{|t|}\frac{1}{P(\cos\phi )}\sum_{n=0}^3 \left[ c_n^{\rm BH} \cos (n\phi ) + s_n^{\rm BH} \sin (n\phi ) \right] \, , \nn \\
|\mathcal{M}_{\textrm{DVCS}} |^2 & \propto & \sum_{n=0}^3 \left[ c_n^{\rm DVCS} \cos (n\phi ) + s_n^{\rm DVCS} \sin (n\phi ) \right]  \, , \nn \\
\mathcal{M}_{\textrm{I}} & \propto & \frac{1}{|t|}\frac{1}{P(\cos\phi )}\sum_{n=0}^3 \left[ c_n^{\rm I} \cos (n\phi ) + s_n^{\rm I} \sin (n\phi ) \right]  \, ,
\label{eq-cross-section} 
\end{eqnarray}
\noindent where $P(\cos\phi )$ comes from the BH lepton propagators. Although  
there are only harmonics up to the maximal order 3 in the sums, the additional 
$\cos\phi$ dependence from the lepton propagators generates in principle an infinite 
series of harmonics for the BH and interference terms. A more detailed harmonic structure
taking into account beam and target polarizations, can be found for instance in 
\ci{Diehl:2005pc}. For transverse target polarization the harmonic series also depends
on the angle $\phi_S$ which specifies the orientation of the target spin vector.
An harmonic analysis of $lp\to lp\gamma$ allows for an examination of the GPDs 
\cite{Diehl:1997bu}. Detailed analytic expressions describing this harmonic structure 
were published in Ref.~\cite{belitsky01}. They involve CFFs, which are integrals of GPDs 
over the momentum fraction $x$ with a hard scattering kernel. At LO the CFFs read~:
\be
{\mathcal F}(\xi,t)\= \int_{-1}^1 dx \Big[e_u^2 F^u+ e_d^2 F^d + e_s^2 F^s\Big]
                       \,\Big[\frac1{\xi-x-i\veps}-\epsilon_f\,
                       \frac1{\xi+x-i\veps}\Big]\,.
\label{eq:CFF-def}
\ee
where $\epsilon_f$ is defined after Eq.\ \req{eq:val-sea}. The CFFs, the
analogues of the convolutions \req{eq:convolutions-mesons}  for DVMP, are
complex functions due to the singularity at $x = \pm \xi$ in the integration
domain.

In Ref.\ \cite{belitsky01} the electroproduction of photons was evaluated to leading and subleading
order in an $1 / Q$ expansion. Since a 
great wealth of data in the valence region \cite{MunozCamacho:2006hx,Girod:2007jq} 
involve not-so-large values of $Q^2$, the impact of this approximation is not
negligible when comparing theoretical expectations and measurements. In 2008
Guichon and Vanderhaeghen \cite{Guichon:2008} elaborated on their previous
numerical computations \cite{VGG, Guichon:1998xv, Vanderhaeghen:1998uc} of the 
$ep \rightarrow ep\gamma$ process to establish analytic expressions of the 
cross section for all polarizations of the proton target and of the incoming
lepton. In these formulas the leptonic tensor is treated exactly,
\textit{i.e.} the involved kinematic terms are kept with
their full $Q^2$ dependence~: they are not expanded as power in $1/Q$. These 
analytic expressions were implemented into a ROOT/C++ code \cite{moutarde09}
and it was checked  in Ref.~\cite{Guidal:2009aa} that they are completely
equivalent to the expressions of cross sections for all polarizations of the proton target
and of the incoming lepton, presented in Refs.~\cite{VGG, Guichon:1998xv, Vanderhaeghen:1998uc}. Later 
Belitsky and Mueller extended their earlier work~\cite{belitsky01} by removing the approximations done
in the $1/Q$ expansion of the leptonic tensor~\cite{Belitsky:2008bz,Belitsky:2010jw}. Note that the case of the transverse target polarization is not treated in these references.

In the present work, we use the theoretical framework of Guichon and Vanderhaeghen
which provides a complete and accurate set of formulas that encompass all types of polarizations, including the case of a transverse target. Our approach is consistent with the so-called BM formalism~\cite{Belitsky:2010jw}.

Further improvements of this approach have been considered: the next-to-leading order (NLO) kernels have been calculated in
Refs.~\ci{ji98,man-piller97,bel97,bel99,Pire:2011st}.  Soft-collinear resummation formulas have been derived recently \cite{Altinoluk:2012fb}. Finite-$t$ and target-mass corrections to DVCS have recently been investigated in the OPE framework in Refs.~\cite{Braun:2012bg,Braun:2012hq} (and references therein). Twist-3 effects in DVCS have also been studied in Refs.~\cite{Anikin:2000em,Radyushkin:2000ap,Kivel:2000fg}.
The phenomenological implications of all these corrections for DVCS observables remain to be studied. This is beyond the scope of the present article.

\section{Comparison to DVCS data}
\label{sec:3}

The $l p \to l p \gamma$ cross section on an unpolarized target for a given beam charge, 
$e_l$ in units of the positron charge and beam helicity $h_l/2$
can be written as~:
\be
d\sigma^{h_l,e_l}(\phi) \= d\sigma_{\rm UU}(\phi)\left[1 + h_l A_{\rm LU,DVCS}(\phi)
                      + e_lh_l A_{\rm LU,I}(\phi) + e_l A_{\rm C}(\phi)\right]\,,
\label{eq-airapetian-asymmetries}
\ee

\noindent where only the $\phi$ dependence of the observables is shown. If both 
longitudinally polarized positively and negatively charged beams are available, 
the asymmetries in Eq.~\req{eq-airapetian-asymmetries} can be isolated, as is the 
case for a large part of HERMES data. Thus, for instance the beam charge asymmetry 
is obtained from the combination~:
\be
A_{\rm C}(\phi) \=\frac1{4d\sigma_{\rm UU}(\phi)}\,\left[   
             ( d\sigma^{\stackrel{+}{\rightarrow}} + d\sigma^{\stackrel{+}{\leftarrow}} )
           - ( d\sigma^{\stackrel{-}{\rightarrow}} 
           + d\sigma^{\stackrel{-}{\leftarrow}}) \right]\,.
\label{eq:A_C}
\ee

 From analogous combinations, one obtains the two beam spin asymmetries $A_{LU,I}$ and 
$A_{LU,DVCS}$:
\begin{eqnarray}
A_{\rm LU,I}(\phi) &\= &\frac1{4d\sigma_{\rm UU}(\phi)}\,\left[   
             (d\sigma^{\stackrel{+}{\rightarrow}} - d\sigma^{\stackrel{+}{\leftarrow}})
            - (d\sigma^{\stackrel{-}{\rightarrow}} - d\sigma^{\stackrel{-}{\leftarrow}}) \right]\,,  \\
A_{\rm LU,DVCS}(\phi) &\= &\frac1{4d\sigma_{\rm UU}(\phi)}\,\left[   
             (d\sigma^{\stackrel{+}{\rightarrow}} - d\sigma^{\stackrel{+}{\leftarrow}})
            + (d\sigma^{\stackrel{-}{\rightarrow}} - d\sigma^{\stackrel{-}{\leftarrow}}) \right]\, .
\end{eqnarray}

If an experiment only has access
to one value of $e_l$ such as in Jefferson Lab, the asymmetries defined in
Eq.~\req{eq-airapetian-asymmetries} cannot be isolated and one can only measure
the beam spin asymmetry $A_{\rm LU}^{e_l}$ which depends on the charge-spin cross section as follows~:
\be
A_{\rm LU}^{e_l}(\phi)\=\frac{d\sigma^{\stackrel{e_l}{\rightarrow}} 
     - d\sigma^{\stackrel{e_l}{\leftarrow}}} {d\sigma^{\stackrel{e_l}{\rightarrow}}
      + d\sigma^{\stackrel{e_l}{\leftarrow}}} \, ,
\ee

\noindent where we use the familiar notation of labeling the charge-spin cross section by the sign of the beam charge $e_l$
and an arrow $\rightarrow$ ($\leftarrow$) for the helicity plus (minus).
One can check that $A_{\rm LU}^{e_l}$ can be written as a function of the spin and 
charge asymmetries defined in Eq.~\req{eq-airapetian-asymmetries}~:
\be
A_{\rm LU}^{e_l}(\phi)\=\frac{e_l A_{\rm LU,I}(\phi)+A_{\rm LU,DVCS}(\phi)}{1+e_lA_{\rm C}(\phi)} \, .
\label{eq-alu-alui-aludvcs}
\ee

The case of longitudinally polarized target observables is simpler, due to the fact that 
there are no data with both varying longitudinal target polarization and beam charge. 
Therefore, experiments measured the target longitudinal spin asymmetry which reads~:
\be
A_{\rm UL}^{e_l}(\phi)\=\frac{ [ d\sigma^{\stackrel{e_l}{\leftarrow\Rightarrow}} 
     + d\sigma^{\stackrel{e_l}{\rightarrow\Rightarrow}} ] - [ d\sigma^{\stackrel{e_l}{\leftarrow\Leftarrow}} 
     + d\sigma^{\stackrel{e_l}{\rightarrow\Leftarrow}} ] } { [ d\sigma^{\stackrel{e_l}{\leftarrow\Rightarrow}} 
     + d\sigma^{\stackrel{e_l}{\rightarrow\Rightarrow}} ] + [ d\sigma^{\stackrel{e_l}{\leftarrow\Leftarrow}} 
     + d\sigma^{\stackrel{e_l}{\rightarrow\Leftarrow}} ] } \, ,
\ee

\noindent where the double arrows $\Leftarrow$ ($\Rightarrow$) refer to the target polarization
state parallel (anti-parallel) to the beam momentum. The double longitudinal target spin asymmetry
is defined in a similar fashion~:
\be
A_{\rm LL}^{e_l}(\phi)\=\frac{ [ d\sigma^{\stackrel{e_l}{\rightarrow\Rightarrow}} 
     + d\sigma^{\stackrel{e_l}{\leftarrow\Leftarrow}} ] - [ d\sigma^{\stackrel{e_l}{\leftarrow\Rightarrow}} 
     + d\sigma^{\stackrel{e_l}{\rightarrow\Leftarrow}} ] } { [ d\sigma^{\stackrel{e_l}{\rightarrow\Rightarrow}} 
     + d\sigma^{\stackrel{e_l}{\leftarrow\Leftarrow}} ] + [ d\sigma^{\stackrel{e_l}{\leftarrow\Rightarrow}} 
     + d\sigma^{\stackrel{e_l}{\rightarrow\Leftarrow}} ] } \, ,
\ee

The HERMES collaboration also had access to a transversally polarized target with both electrons and positrons. They therefore
were able to measure two types of observables~:
\begin{eqnarray}
A_{\rm UT,I}(\phi,\phi_S) & & \=\nn \\
& & \frac{ d\sigma^+(\phi,\phi_S)- d\sigma^+(\phi,\phi_S+\pi)+d\sigma^-(\phi,\phi_S)- d\sigma^-(\phi,\phi_S+\pi) } 
{ d\sigma^+(\phi,\phi_S)- d\sigma^+(\phi,\phi_S+\pi)+d\sigma^-(\phi,\phi_S)- d\sigma^-(\phi,\phi_S+\pi) } \,,  \\
A_{\rm UT,DVCS}(\phi,\phi_S) & & \=\nn \\
& & \frac{ d\sigma^+(\phi,\phi_S)- d\sigma^+(\phi,\phi_S+\pi)-d\sigma^-(\phi,\phi_S)+ d\sigma^-(\phi,\phi_S+\pi) } 
{ d\sigma^+(\phi,\phi_S)- d\sigma^+(\phi,\phi_S+\pi)+d\sigma^-(\phi,\phi_S)- d\sigma^-(\phi,\phi_S+\pi) } \, .
\label{eq:A_UTDVCS}
\end{eqnarray}

Finally, it is worth mentioning that the HERMES collaboration usually does not publish the 
$\phi$-dependence of the asymmetries but rather chose to extract harmonics out of their 
asymmetries. For instance, in the case of the beam charge asymmetry $A_C$, the $\cos(n\phi )$
harmonics are extracted using the following formula~:
\be
A_C^{\cos(n\phi )} = N \int_0^{2\pi} d\phi A_C(\phi )\cos(n\phi ) \,,
\label{eq:projection}
\ee

\noindent where the normalization factor $N$ is $1/2\pi$ in the case $n=0$ and $1/\pi$ for 
$n\geq 1$. Since HERMES has not measured numerators ($\mathcal D$) and denominators 
($\mathcal S$) of the asymmetries in Eqs.~\req{eq:A_C} - \req{eq:A_UTDVCS} separately the 
projection \req{eq:projection} is the best approximation to the Fourier coefficients
\be
          \frac12 \frac{\int_0^{2\pi} d\phi {\mathcal D}(\phi) \cos(n\phi)}
                       {\int_0^{2\pi} d\phi {\mathcal S}(\phi)}  
\ee
HERMES can do. The latter Fourier coefficients are closest related to the CFFs.

The coefficients of the various Fourier harmonics occurring in Eq.~\req{eq-cross-section}
provide information about CFFs, or equivalently GPDs, either in the interference with the real
Bethe-Heitler amplitude or from the DVCS process. Through the measurement of cross sections 
or the various asymmetries described in Eqs.~\req{eq:A_C}--\req{eq:A_UTDVCS}, one can put 
constraints on different combination of GPDs. As an illustration we quote the leading-twist, 
LO pQCD connection between observables and CFF for some 
asymmetries~\ci{belitsky01,Belitsky:2010jw,Diehl:2005pc}~:
\ba
A_C^{\cos{\phi}} &\propto& \mbox{Re}\, \Big[F_1 \mathcal{H} +\xi (F_1+F_2)\mathcal{\widetilde{H}}
                    -\frac{t}{4m^2} F_2 \mathcal{E}\Big]\,, \nn\\
A^{\sin{\phi}}_{LU,I}&\propto& \mbox{Im}\, \Big[F_1 \mathcal{H} 
                     +\xi (F_1+F_2)\mathcal{\widetilde H}
                    -\frac{t}{4m^2} F_2 \mathcal{E}\Big]\,, \nn\\
A^{\sin{\phi}}_{UL,I} &\propto& \mbox{Im}\,\Big[\xi(F_1+F_2)(\mathcal{H}
                  +\frac{\xi}{1+\xi}\mathcal{E}) + F_1\widetilde{\mathcal{H}}-
      \xi(\frac{\xi}{1+\xi}F_1+\frac{t}{4M^2}F_2)\widetilde{\mathcal{E}}\Big]\,,
      \nn \\
A_{LL,I}^{\cos{\phi}} &\propto& \mbox{Re}\, \Big[\xi(F_1+F_2)(\mathcal{H}
                  +\frac{\xi}{1+\xi}\mathcal{E}) + F_1\widetilde{\mathcal{H}}-
      \xi(\frac{\xi}{1+\xi}F_1+\frac{t}{4M^2}F_2)\widetilde{\mathcal{E}}\Big]\,,
      \nn \\
A_{LL,DVCS}^{\cos{(0\phi)}} &\propto&  \mbox{Re}\,
     \Big[4(1-\xi^2)\big(\mathcal{H}\widetilde{\mathcal{H}}^*
                  +\widetilde{\mathcal{H}}\mathcal{H}^*\big)
              -4\xi^2\big(\mathcal{H}\widetilde{\mathcal{E}}^*
              + \widetilde{\mathcal{E}} \mathcal{H}^* 
              + \widetilde{\mathcal{H}} \mathcal{E}^*
              + \mathcal{E}  \widetilde{\mathcal{H}}^*\big) \nn\\
              &&\qquad -4\xi\big(\frac{\xi^2}{1+\xi} + \frac{t}{4M^2}\big)\,
                \big(\mathcal{E}\widetilde{\mathcal{E}}^* 
              + \widetilde{\mathcal{E}}\mathcal{E}^*\big)\Big]\,, \nn\\
A_{UT,DVCS}^{\sin{(\phi-\phi_s)}} &\propto& 
                        \Big[\mbox{Im}\,(\mathcal{H}\mathcal{E}^*)
       -\xi\mbox{Im}\,(\widetilde{\mathcal{H}}\widetilde{\mathcal{E}}^*)\Big]\,,
                        \nn\\
A_{UT,I}^{\sin{(\phi-\phi_s)}\cos{\phi}} &\propto& \mbox{Im}\,
              \Big[-\frac{t}{4M^2}\big(F_2\mathcal{H}-F_1\mathcal{E}\big)
     +\xi^2\big(F_1+\frac{t}{4M^2}F_2\big)\big(\mathcal{H}+\mathcal{E}\big)\nn\\
   &&\qquad  -\xi^2\big(F_1+F_2)\big(\widetilde{\mathcal{H}} 
              + \frac{t}{4M^2}\widetilde{\mathcal{E}}\big)\Big]\,.
\label{eq:asymmetries}
\ea

\begin{table}[p]
\begin{center} 
    \begin{tabular}{|c|c|c|}
        \hline
        \rule[-3mm]{0mm}{8mm}\bf Experiment   & \bf Observable                                         & \bf Normalized CFF dependence \\ \hline\hline
          ~                                   	& \rule[-2mm]{0mm}{7mm}$A_{\rm C}^{\cos 0\phi}$                    		& ${\rm Re} \mathcal H +0.06 {\rm Re} \mathcal E +0.24 {\rm Re} {\widetilde{\mathcal H}} $              \\ \cline{2-3}
         ~ 	& \rule[-2mm]{0mm}{7mm}$A_{\rm C}^{\cos \phi}$                      		& ${\rm Re} \mathcal H +0.05 {\rm Re} \mathcal E +0.15 {\rm Re} {\widetilde{\mathcal H}} $              \\ \cline{2-3}
         ~     	& \rule[-2mm]{0mm}{7mm}$A_{\rm LU,I}^{\sin \phi}$                   		& ${\rm Im} \mathcal H +0.05 {\rm Im} \mathcal E +0.12 {\rm Im} {\widetilde{\mathcal H}} $              \\ \cline{2-3}
         ~   	& \rule[-2mm]{0mm}{7mm}$A_{\rm UL}^{+,\sin \phi}$                     		& ${\rm Im} {\widetilde{\mathcal H}} + 0.10{\rm Im} \mathcal H + 0.01{\rm Im} \mathcal E$              \\ \cline{2-3}
        HERMES     	& \rule[-2mm]{0mm}{7mm}$A_{\rm UL}^{+,\sin 2\phi}$                    		& ${\rm Im} {\widetilde{\mathcal H}} - 0.97{\rm Im} \mathcal H    + 0.49{\rm Im} \mathcal E  - 0.03{\rm Im}  {\widetilde{\mathcal E}}$           \\ \cline{2-3}
        ~     	& \rule[-2mm]{0mm}{7mm}$A_{\rm LL}^{+,\cos 0\phi}$                    		& $1+ 0.05{\rm Re} {\widetilde{\mathcal H}} + 0.01{\rm Re} \mathcal H$           \\ \cline{2-3}
        ~     	& \rule[-2mm]{0mm}{7mm}$A_{\rm LL}^{+,\cos \phi}$                    		& $1+0.79{\rm Re} {\widetilde{\mathcal H}} +0.11{\rm Im} \mathcal H$           \\ \cline{2-3}
         ~   	& \rule[-2mm]{0mm}{7mm}$A_{\rm UT,DVCS}^{\sin (\phi -\phi_S)}$      	&  ${\rm Im} {\mathcal H} {\rm Re} {\mathcal E} - {\rm Im} {\mathcal E} {\rm Re} {\mathcal H} $          \\ \cline{2-3}
        ~    	& \rule[-2mm]{0mm}{7mm}$A_{\rm UT,I}^{\sin (\phi -\phi_S)\cos\phi}$ 	& ${\rm Im} \mathcal H -0.56 {\rm Im} \mathcal E - 0.12 {\rm Im} {\widetilde{\mathcal H}}$              \\ \hline\hline
        ~   					& \rule[-2mm]{0mm}{7mm}$A_{\rm LU}^{-,\sin \phi}$                          		& ${\rm Im} {\mathcal H} + 0.06 {\rm Im} \mathcal E + 0.21 {\rm Im} {\widetilde{\mathcal H}}$         \\ \cline{2-3}
        CLAS      				& \rule[-2mm]{0mm}{7mm}$A_{\rm UL}^{-,\sin \phi}$                     		& ${\rm Im} {\widetilde{\mathcal H}}  + 0.12{\rm Im} \mathcal H + 0.04{\rm Im} \mathcal E$             \\ \cline{2-3}
        ~      					& \rule[-2mm]{0mm}{7mm}$A_{\rm UL}^{-,\sin 2\phi}$                    		& ${\rm Im} {\widetilde{\mathcal H}} - 0.79{\rm Im} \mathcal H + 0.30{\rm Im} \mathcal E - 0.05{\rm Im}  {\widetilde{\mathcal E}}$              \\ \hline\hline
        ~ 					& \rule[-2mm]{0mm}{7mm}$\Delta\sigma^{\sin\phi}$                       		& ${\rm Im} {\mathcal H} + 0.07 {\rm Im} \mathcal E + 0.47 {\rm Im} {\widetilde{\mathcal H}} $              \\ \cline{2-3}
        {HALL A} & \rule[-2mm]{0mm}{7mm}$\sigma^{\cos 0\phi}$  &  $1+0.05{\rm Re} \mathcal H +0.007\mathcal H \mathcal
        H^*$          \\ \cline{2-3}
        ~ & \rule[-2mm]{0mm}{7mm}$\sigma^{\cos\phi}$  &  $1+0.12{\rm Re} \mathcal H +0.05 {\rm Re} {\widetilde{\mathcal H}} $          \\ \hline\hline
        HERA   				& \rule[-2mm]{0mm}{7mm}$\sigma_{\rm
          DVCS}$                                     	& $\mathcal H \mathcal
        H^* +0.09 \mathcal E \mathcal E^*  + {\widetilde{\mathcal
            H}}{\widetilde{\mathcal H}}^*    $         \\
        \hline
    \end{tabular}
\caption{\small Dependence of the observables on the CFF at the kinematics specified in 
Tab.~\ref{table:AvgKin}. The coefficients in front of the CFF are normalized to the largest 
one, but only relative coefficients larger than 1\% are kept, except for the Hall~A 
cross section, where we also show the term quadratic to the CFF $\mathcal H$ since it 
contributes significantly. 
In order to get simple CFF dependences for this table, the unpolarized
cross section in the denominator of the asymmetries is approximated by the
term $c_0^{\rm BH}/P(\cos{\phi})$ in Eq.~\req{eq-cross-section}.
}
\label{table:CFFdep}
\end{center} 
\end{table}

\begin{table}[p]
\begin{center} 
    \begin{tabular}{|c|c|c|c|}
\hline
~  & \multicolumn{3}{|c|}{\bf Kinematics} \\ \cline{2-4}
\raisebox{1.5ex}[0pt]{\bf Experiment}		  & \rule[-2mm]{0mm}{7mm}$x_B$	&	$Q^2$ [GeV$^2$]	&	  $t$ [GeV$^2$]  \\ \hline \hline
HERMES	&0.09	& 2.50	&-0.12 \\ \hline
CLAS		&0.19	& 1.25	&-0.19 \\ \hline
HALL A	&0.36	& 2.30	&-0.23 \\ \hline
HERA		&0.001	& 8.00	&-0.30 \\
        \hline
    \end{tabular}
\caption{\small Typical kinematics used in Tab.\ \ref{table:CFFdep} for various experiments. }
\label{table:AvgKin}
\end{center} 
\end{table}

In the following subsections, we compare predictions on DVCS evaluated from the
GPDs described in Sect.\ \ref{subsec:2.1} to the available data in different 
kinematic domains~: from small $\xbj$ (H1, ZEUS) \cite{Aktas:2005ty}--\ci{zeus-dvcs} 
to intermediate $\xbj$ (HERMES) \ci{Airapetian:2008aa}--\ci{Airapetian:2012pg}
and large $\xbj$ (Jefferson Lab Hall~A and CLAS) 
\cite{MunozCamacho:2006hx,Girod:2007jq,Chen:2006na}.
The HERA H1 and ZEUS collaborations measured the DVCS ($\gamma^*p\to \gamma p$) cross 
section at large $W$ and $Q^2$ but small $\xbj$, which are mostly sensitive to the 
imaginary part of $\mathcal H$. HERMES measured and published a large number of 
asymmetries with different beam charge, helicity states and target polarization. These 
data lie in the range $0.05 \lesssim \xbj \lesssim 0.25$ and are sensitive to the real 
and imaginary parts of the CFFs $\mathcal{H}$, $\mathcal{E}$ and $\widetilde{\mathcal{H}}$. 
The data sets from Jefferson Lab either cover a wide kinematic range ($0.11 \lesssim
\xbj \lesssim 0.58$) \cite{Girod:2007jq} or are  highly precise on a
restricted kinematic domain \cite{MunozCamacho:2006hx}. In the first case beam
spin asymmetries are measured, and in the second helicity-dependent and
independent cross sections. The dependence of all these observables to the CFFs are 
compiled in Tab.\ \ref{table:CFFdep} where for each experiment, we have chosen the 
typical kinematics listed in Tab.\ \ref{table:AvgKin}.
The BH amplitude is evaluated from the Kelly parametrization of nucleon form factors 
\cite{Kelly:2004hm}.
We would like to stress once again that no DVCS data has been used in order to fix the
GPD parameters. Observables which vanish to the accuracy we are working and for which the
data are compatible with zero within errors will not be discussed in the
following. The propagation of PDF errors to the GPDs $H$ and ${\widetilde H}$
is the only source of uncertainties we have considered so far. They will be 
shown as shadowed bands around the predictions on all the following figures.

\subsection{Beam charge asymmetry}
\label{subsec:3.1}
\begin{figure}[t]
\begin{center}
\includegraphics[width=0.6\tw]{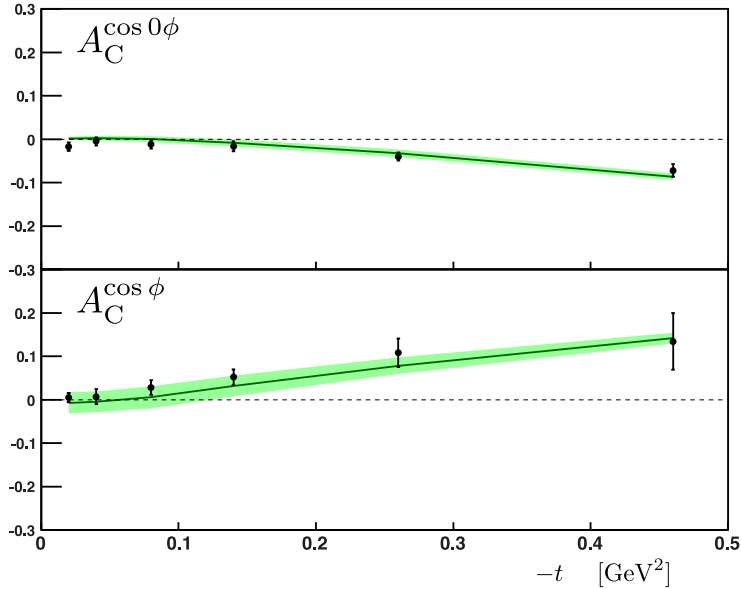}
\caption{The $\cos 0\phi$ and $\cos\phi$ harmonics of the beam charge
asymmetry at the kinematical setting $x_B\simeq 0.097$ and $Q^2\simeq
2.51\, \gev^2$. Data are taken from HERMES \ci{Airapetian:2012mq} --Tab.~6.
Our results, shown as solid lines with the shaded areas as
the error bands, are evaluated at the kinematics specified
in Ref.~\ci{Airapetian:2012mq} --Tab.~6, and joined by straight lines to guide the eyes.} 
\label{fig:hermes-BCA}
\end{center}
\end{figure}

We begin the discussion with the beam charge asymmetry, $A_C$, as measured by the HERMES 
collaboration during 1996--2007, a result which has been recently updated in 
Ref.~\ci{Airapetian:2012mq}. This observable is generated by the BH-DVCS interference term 
(see Eq.~\req{eq:asymmetries}) which, to leading-twist accuracy, feeds the
$\cos{n\phi}$ harmonics ($n=0,1$). As inspection of Tab.\ \ref{table:CFFdep} reveals,  
both the harmonics depend mostly on the real part of the CFF $\mathcal H$.
In Fig.~\ref{fig:hermes-BCA}, our predictions are compared to the HERMES data
\ci{Airapetian:2012mq}. Very good agreement can be seen for both the
harmonics, demonstrating that in this kinematical range, the real part of
$\mathcal H$ has the right magnitude.

\subsection{Beam spin asymmetry}
\label{subsec:3.2}

As can be seen from Eq.~\req{eq:asymmetries} the $\sin{\phi}$ harmonic of the beam 
spin asymmetry, $A_{LU,I}$, depends on the same combination of electromagnetic 
form factors and CFFs as the beam charge asymmetry. 
As we already mentioned this combination is dominated by the GPD $H$, see Tab.\ 
\ref{table:CFFdep}. Since the electromagnetic form factors are real, the imaginary 
part of the CFF ${\mathcal H}$ is required which, to leading-order of perturbative 
QCD (see \req{eq:CFF-def}), is given by the GPD at the cross-over line $x=\xi$, 
{\it i.e.}\ $A^{\sin{\phi}}_{\rm LU,I}$ essentially probes the combination
\be
e_u^2H^u(\xi,\xi,t)+e_d^2H^d(\xi,\xi,t) +e_s^2H^s(\xi,\xi,t)\,.
\ee

\begin{figure}[htbp]
\begin{center}
\includegraphics[width=0.7\tw]{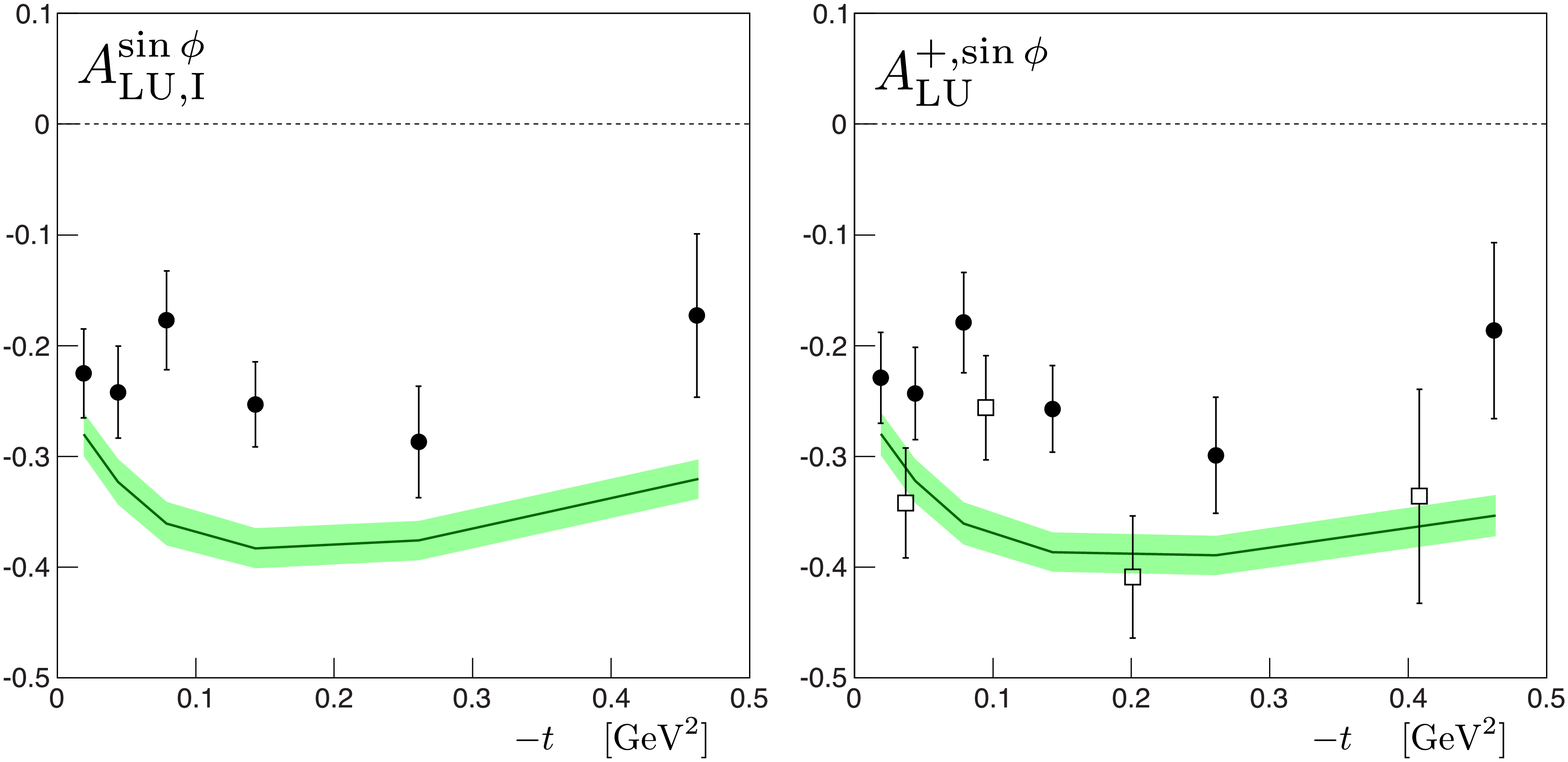}
\caption{Left plot: $A_{\rm LU,I}^{\sin\phi}$ as a function of $-t$ measured
  by the HERMES collaboration \ci{Airapetian:2012mq}--Tab.~5. Right plot: 
  $A_{\rm LU}^{+,\sin\phi}$ versus $-t$ obtained from the non-recoil data on 
$A_{\rm LU,I}^{\sin\phi}$ and $A_{\rm C}^{\cos 0\phi}$ measured by the HERMES collaboration 
\ci{Airapetian:2012mq} --Tab.~5 and 6 (solid circles, see text for details) and the 
more recent recoil data \ci{Airapetian:2012pg} (open squares). 
For other notations and the values of the averaged kinematic variables,
refer to Fig.~\ref{fig:hermes-BCA}.}
\label{fig:hermes-BSA}
\end{center}
\end{figure}

Our results for the beam spin asymmetry $A_{LU,I}^{\sin\phi}$ are shown in Fig.\ 
\ref{fig:hermes-BSA}-left and compared to the HERMES data \ci{Airapetian:2012mq}. The 
agreement between predictions and data is not as good in this case, our results 
differ by about $40\%$ ($\simeq 0.1$ in  absolute value) from experiment.
Recently the HERMES collaboration has published data on the $\sin \phi$ harmonic of 
the beam spin asymmetry using a recoil detector and a positron beam 
\ci{Airapetian:2012pg}. In this experiment all three final state particles are detected 
and therefore the resonant background severely reduced. In so far the recoil data are
closer to the exclusive process $l p\to l p \gamma$ to which our theory applies. 
The data were taken at about 
the same average values  of $x_B$ and $Q^2$. In order to compare to recoil data, we 
computed $A_{LU}^+$ using Eq.\ \req{eq-alu-alui-aludvcs} with $A_{LU, I}$ and $A_C$ from 
the non-recoil data and $A_{LU, DVCS} = 0$ (exact at twist 2 and in agreement with experimental results from Ref.~\ci{Airapetian:2012mq}).
Then the $\sin \phi$ coefficient is~:
\begin{equation}
A_{LU}^{+\sin \phi} \simeq \frac{A_{LU, I}^{\sin \phi}}{1+A_C^{\cos 0 \phi}}
\end{equation}
On the right hand side of Fig.\ \ref{fig:hermes-BSA}  we therefore show both
$A_{LU}^{+\sin \phi}$  from the non-recoil and the recoil data. We observe that 
the recoil data are significantly larger in absolute value, yielding very good 
agreement with our predictions. Similar effects for other DVCS observables may occur 
but with the exception of the beam spin asymmetry, there are no measurements with the
recoil detector available. The effect of the resonant background in other 
observables is unknown. Note that $A_{LU}^{\sin{\phi}}$ vanishes for forward scattering,
$t=t_{\rm min}$. The trend towards zero is however only visible for $t$ of order 
$t_{\rm min}=-4m^2 \xi^2/(1-\xi^2)$ which is very small, about $-0.02\,\gev^2$ for 
HERMES kinematics.

\begin{figure}[htbp]
\begin{center}
\includegraphics[width=1\tw,]{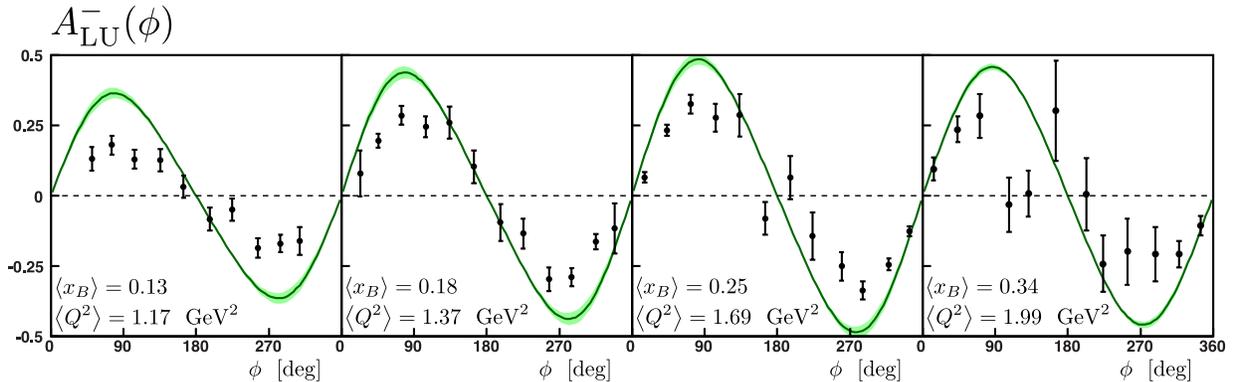}
\caption{The beam spin asymmetry $A_{\rm LU}(\phi )$ measured by the CLAS
  collaboration \ci{Girod:2007jq} for a sample of 4 bins, all taken at an
  average $-t$ value of $0.3$~GeV$^2$. For other notations refer to Fig.\ 
\ref{fig:hermes-BCA}.}
\label{fig:clas-LU}
\end{center}
\end{figure}

The CLAS collaboration published accurate data on the beam helicity asymmetry in a
large kinematical range \ci{Girod:2007jq}. The data analysis required the detection of
the full $(e,p,\gamma )$ final state, much like the HERMES recoil data.
A selected sample of the $\phi$-dependent
asymmetry $A^-_{\rm LU}(\phi )$ is shown on Fig.~\ref{fig:clas-LU}. While the dominance 
of the $\sin{\phi}$ term is clearly visible in the data one notices a 
discrepancy between the data and our prediction of the same order as for the non-recoil
HERMES beam spin asymmetry, see above. We repeat - our GPDs  are optimized for small 
skewness (see discussion in Sect.~\ref{subsec:2.1} and Sect.~\ref{subsec:2.2}) and 
one therefore cannot expect perfect agreement with large-skewness experiment.  

\subsection{Longitudinal target polarization}
\label{subsec:3.3}
 
\begin{figure}[t]
\begin{center}
\includegraphics[width=.7\tw]{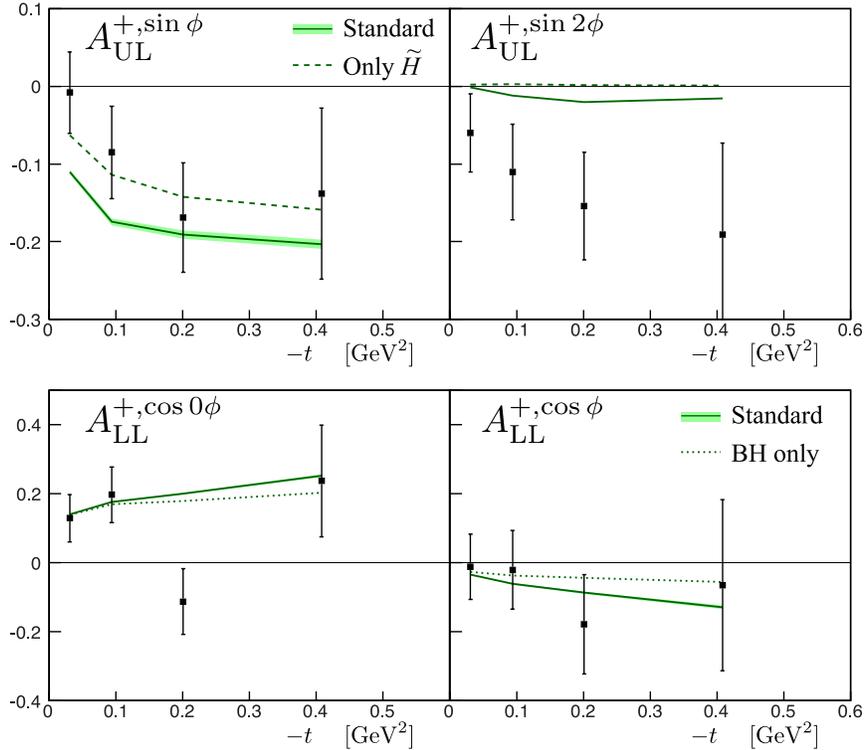}
\caption{Top plots: The $\sin{\phi}$ and the $\sin{2\phi}$ harmonics of
  $A_{\rm UL}^+$. Bottom plots: The $\cos{0\phi}$ and the $\cos{\phi}$ harmonics
  of $A_{\rm LL}^+$. HERMES data are taken from Ref.~\ci{Airapetian:2010ab}, the average kinematics
is $x_B=0.1$ and $Q^2$=2.46~GeV$^2$.
The dashed lines in the top plots represent the results with only the
contributions of $\widetilde{\mathcal{H}}$ and the dotted lines in the bottom
plots the BH contributions.
Our results shown as solid lines with the shaded areas as the error bands,
are evaluated at the specified kinematics corresponding to each bin in $-t$.
For other notations refer to Fig.\ \ref{fig:hermes-BCA}.} 
\label{fig:hermes-L}
\end{center}
\end{figure}

This asymmetry is very similar to the beam spin asymmetry in that it is 
dominated by the DVCS-BH interference. The most important observable of this 
type is $A_{\rm UL}^{+\sin{\phi}}$ which is primarily sensitive to
$\widetilde{\mathcal{H}}$ as an inspection of Tab.\ \ref{table:CFFdep}
reveals.  The comparison of the HERMES data \ci{Airapetian:2010ab} on this 
observable measured with a positron beam, with our predictions is made in 
Fig.~\ref{fig:hermes-L} and, given 
the large experimental errors, reasonable agreement is observed.
Only at small $-t$ our result seems slightly larger in absolute value
than the data by a fraction of the HERMES error bar. A surprisingly large 
$\sin{2\phi}$ harmonic appears in the HERMES data, see Fig.\ \ref{fig:hermes-L}. 
Theoretically this contribution should be heavily suppressed to the order we 
are working and indeed our prediction is very small. The reason for the large 
experimental value of $A_{UL}^{+\sin{2\phi}}$ is unclear. For both $A_{UL}^+$ 
parameters the figure also shows the contribution from $\widetilde H$ separately, 
confirming that this GPD represents a significant fraction of $A_{UL}^{+\sin{\phi}}$.

\begin{figure}[t]
\begin{center}
\includegraphics[width=0.7\tw]{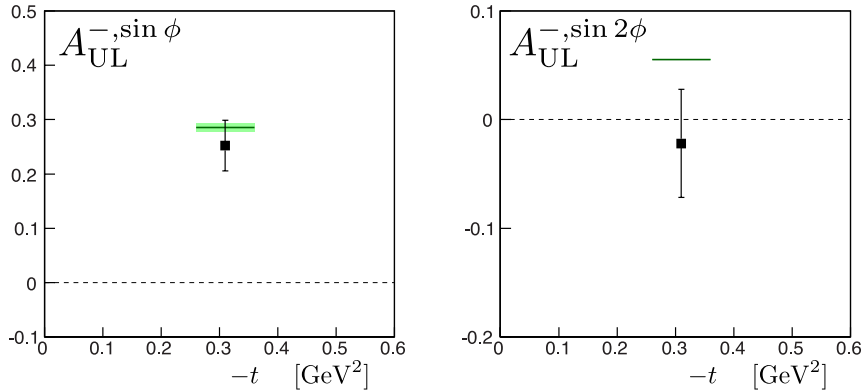}
\caption{CLAS longitudinal target single spin asymmetry $A_{UL}^-$ \ci{Chen:2006na} 
  at the average kinematics $x_B=0.28$, $-t=0.31$~GeV$^2$ and $Q^2=1.82$~GeV$^2$. 
  The left plot shows the $\sin\phi$ moment versus $-t$ whereas the right plot shows 
  the $\sin 2\phi$ moment. In both plots, our results are shown as solid lines 
  with the shaded areas as the error bands.}
\label{fig:clas-L}
\end{center}
\end{figure}

The CLAS collaboration has published data on $A_{UL}^-$ \ci{Chen:2006na},
integrated over a large kinematical bin. In contrast to HERMES this collaboration 
uses an electron beam implying that the BH-DVCS interference contributes with opposite
sign to $A_{UL}^-$ than in the HERMES data on $A_{UL}^+$. The $\sin\phi$ and $\sin 2\phi$
harmonics are shown in Fig.~\ref{fig:clas-L} along with our predictions.
In contrast to the HERMES data, the  $\sin 2\phi$ harmonic is compatible with 
zero, however our prediction is small and positive, off by about 1.5~$\sigma$.
For the $\sin\phi$ harmonic, our result and the CLAS data are in perfect agreement.

HERMES has also measured the double-spin asymmetry for a longitudinally
polarized beam and target. In contrast to the single spin asymmetries
$A_{UL}^+$ and $A_{LU}^+$ the BH term contributes to $A_{LL}^+$. In fact it provides 
the dominant contribution to the $\cos{0\phi}$ term and about half the
magnitude of the $\cos{\phi}$ harmonic, see Fig.\ \ref{fig:hermes-L}. Thus,
at the current experimental  accuracy in these observables, no information
about GPDs can be extracted. Our full results for $A_{LL}^+$ are of course in 
agreement with HERMES data as can be seen from Fig.\ \ref{fig:hermes-L}.

\subsection{Transversely polarized target}
\label{subsec:3.4}

The $\sin(\phi-\phi_S)$ harmonic of the transverse target spin asymmetry 
is especially interesting since it is generated by the squared DVCS contribution 
and it is mostly sensitive to ${\rm Im}({\mathcal H}{\mathcal E}^*)$, see  
Eq.~\req{eq:asymmetries}  and Tab.~\ref{table:CFFdep}. For given $\mathcal H$ it 
is uniquely sensitive to the CFF $\mathcal E$. It therefore constitutes one of the 
few ways to constrain the GPD $E$. The valence 
part of $E$ is rather well constrained by fits to the data of the nucleon form
factors \ci{DFJK4} and is in fair agreement with the data on $A_{UT}$ for
$\rho^0$ electroproduction \ci{Airapetian:2009ad,Adolph:2012ht}. As we
discussed in Sect.\ \ref{subsec:2.2} for $\rho^0$ production the contributions
from $E$ for sea quarks and gluons cancel to a large extent. In view of this
the DVCS data on $A_{UT}$ are complementary to DVMP and especially important
because to LO QCD there is no contribution from $E^g$ and therefore the
cancellation between the gluon and sea-quark contributions cannot happen.
Hence, $A_{UT}$ for DVCS probes the sea-quark part of $E$ for given GPDs $H$
and $E_{\rm val}$. We remark that $A_{UT,DVCS}^{\sin(\phi-\phi_S)}$ is forced to vanish 
for forward scattering by angular momentum conservation.

\begin{figure}[t]
\begin{center}
\includegraphics[width=0.9\tw]{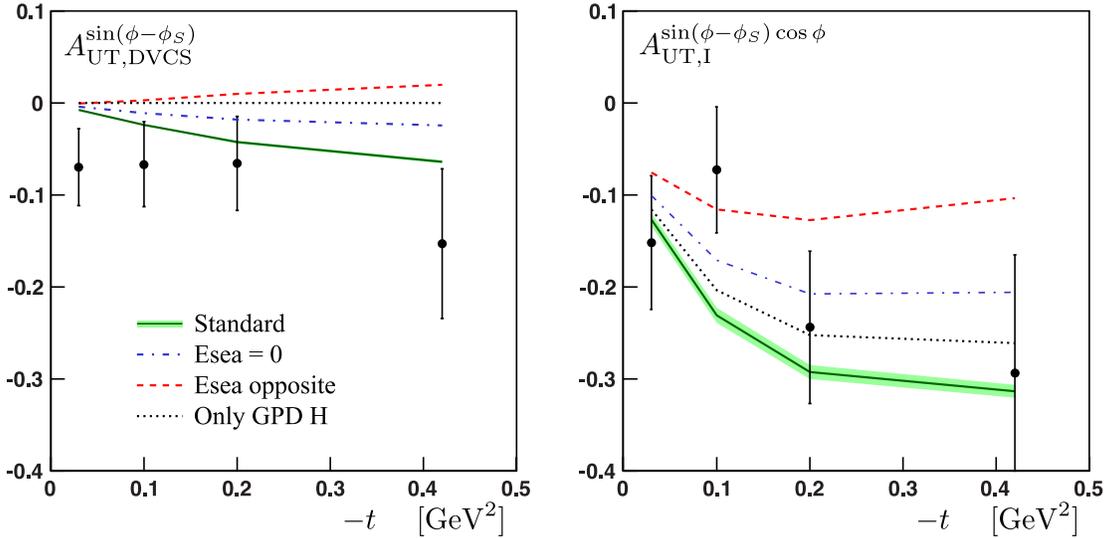}
\caption{The transverse target spin asymmetries. The left plot shows  
  $A^{\sin(\phi - \phi_S)}_{\rm UT,DVCS}$  whereas the right plot shows 
  $A^{\sin (\phi - \phi_S) \cos\phi}_{\rm UT,I}$. Data are taken from
  Ref.~\ci{Airapetian:2008aa}, their averaged kinematics is $x_B=0.09$ and $Q^2=2.5$~GeV$^2$.
  Our results are shown for the standard scenario ($N_s=-0.155$ ) as well as for
  two other normalizations of $E_{\rm sea}$ ($N_s=0$ and 0.155) and finally keeping 
  only GPD $H$ and setting the others to zero. For other notations refer to Fig.\ 
  \ref{fig:hermes-BCA}.}
\label{fig:Hermes_TSAT}
\end{center}
\end{figure}

The HERMES data \ci{Airapetian:2008aa} on the $\sin(\phi-\phi_S)$ harmonics of $A_{UT}$
is shown in Fig.\ \ref{fig:Hermes_TSAT}. Also shown in  Fig.~\ref{fig:Hermes_TSAT}
is $A^{\sin (\phi - \phi_S) \cos\phi}_{\rm  UT,I}$. This observable receives separate 
contributions from Im$\mathcal H$, Im$\mathcal E$ and Im$\widetilde{\mathcal H}$
(see Eq.~\req{eq:asymmetries} and Tab.~\ref{table:CFFdep}) and even for 
$E=\widetilde{H}=0$ reasonable agreement with experiment is achieved. In contrast
to this observable $A^{\sin (\phi - \phi_S)}_{\rm UT,DVCS}$  would vanish if $E$ was zero
which is not the case experimentally. 

In order to check the normalization of the GPD $E$ for a flavor symmetric sea 
we also display in Fig.~\ref{fig:Hermes_TSAT} results for three different
normalizations of $E^s$: $N_s=0$ and $N_s=\pm 0.155$; for the latter value a
positivity bound is saturated \ci{GK4} (see Sect.\ \ref{subsec:2.1.2}). It can
be seen from the figure that a negative $E^s$ seems to be favored. Considering the
large errors of the data, a value of $N_s$ close to zero but negative is not excluded
while a large positive value is apparently in conflict with experiment.

\subsection{Hall~A cross sections}
\label{subsec:3.5}

The helicity-dependent $ep\to ep\gamma$ cross section has been
measured by the Hall~A collaboration at Jefferson Lab \ci{MunozCamacho:2006hx} 
at fixed $x_B$=0.36. These data are extremely accurate and are therefore very
demanding to theory. They have the advantage to cleanly separate the
difference of the cross sections for opposite electron helicities, $\Delta \sigma$,
from their sum, $\Sigma\sigma$. The main contribution to the cross section 
difference is the $\sin \phi$ harmonic fed by the BH-DVCS interference. From 
Tab.~\ref{table:CFFdep} we see that the most prominent contribution from the CFFs is 
${\rm Im}{\mathcal H}$. The cross section sum, on the other hand, which corresponds 
to the unpolarized cross section, receives contributions from all three terms BH,
DVCS and the BH-DVCS interference but the BH contribution is dominant. 
Its most prominent harmonics are $\cos 0\phi$ and $\cos \phi$.  
\begin{figure}[htb]
\begin{center}
\includegraphics[width=1.0\tw]{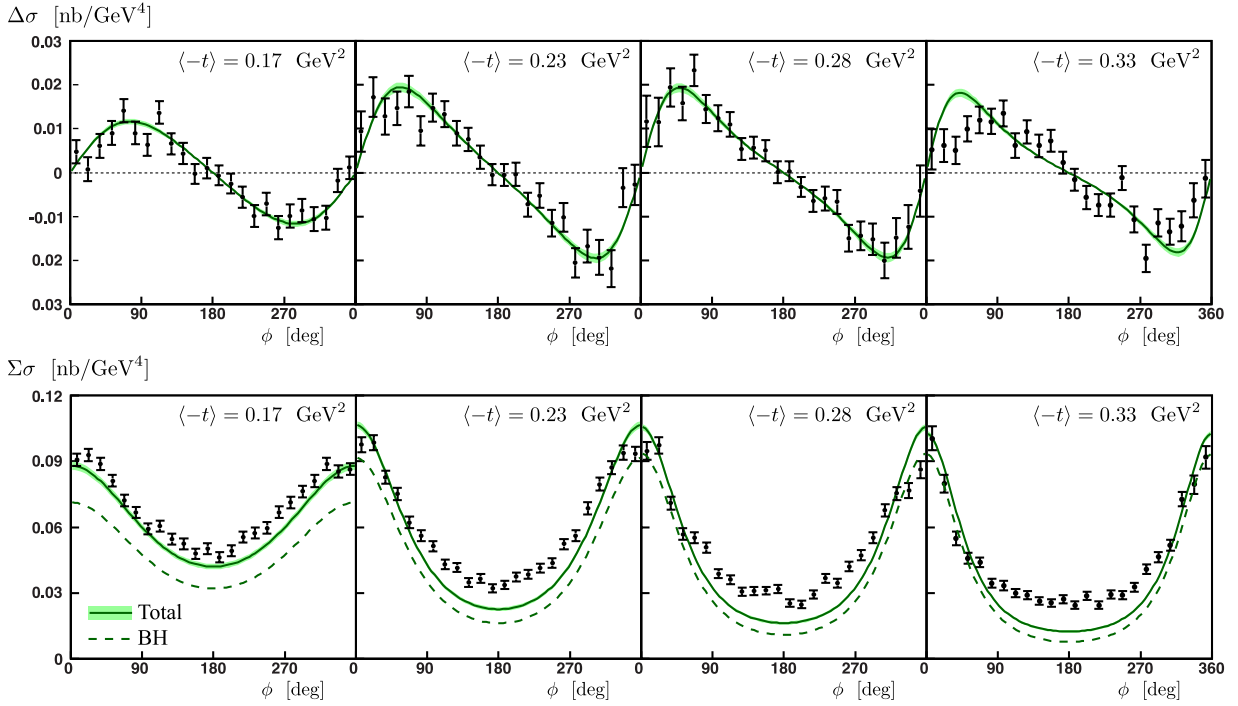}
\caption{Jefferson Lab Hall~A helicity-dependent cross section data at
  and different $t$ bins for $x_B=0.36$ and $Q^2=2.3$~GeV$^2$.
The top plots show the differences of cross
  sections for opposite electron helicities versus $\phi$  whereas the
  bottom plots show the unpolarized cross section. Data are taken from 
  \ci{MunozCamacho:2006hx}.The Bethe-Heitler contribution to the 
  unpolarized cross section is represented by dashed lines whereas our
  full results are shown as solid lines with the errors as shadowed bands.} 
\label{fig:hall-A}
\end{center}
\end{figure}

The comparison of the difference and sum of the Hall~A helicity cross sections
with our results at their highest $Q^2$ bin is shown in Fig.~\ref{fig:hall-A}. 
The agreement is nearly perfect for $\Delta \sigma$ which 
indicates that our parameterization of $H$ seems to be adequate. 
This is in line with our findings for $A_{LU,I}^{\sin\phi}$ at HERMES 
kinematics, namely that our predictions agree well with the recoil
data \ci{Airapetian:2012pg}.

The cross section sum exhibits the  expected $\phi$ dependence of a linear
combination of the $\cos 0\phi$ and $\cos\phi$ harmonics. There is 
however a discrepancy between theory and experiment of up to 30\% in 
the strength, especially around $\phi=180^\circ$, pointing to something missing in
the real parts of the various CFFs entering this observable, in
particular ${\rm Re}{\mathcal H}$, see Tab.\ \ref{table:CFFdep}. Note that by taking the ratio
of the difference over the sum of cross-sections for Hall~A, one recovers a result fully
compatible with the CLAS beam spin asymmetry data in the same kinematical range. It is therefore
no surprise that our approach also has difficulties reproducing the CLAS beam spin
asymmetries since the
issue apparently lies in the unpolarized cross section. The real part of
${\mathcal H}$ is also probed by $A_C^{\cos n\phi}$. Since ${\rm Re}{\mathcal H}$  
contributes to $\Sigma\sigma$ with a small coefficient as compared to the
BH-term, a substantial increase of ${\rm Re}{\mathcal H}$ would be required in 
order to fit the Hall~A $\Sigma\sigma$ data. Whether such an increase is compatible 
with other DVCS and DVMP data remains to be seen. In any case,
we would like to stress again that the GPD parametrization we are using has
been tuned to much lower values of $\xi$. Thus, for instance, the addition of a
D-term to the double-distribution might improve the agreement
between theory and experiment. The exploration of this as well as other
improvements of the GPDs is left for future work.

\subsection{H1 and ZEUS cross sections}
\label{subsec:3.6}

\begin{figure}[p]
\begin{center}
\includegraphics[width=.9\tw]{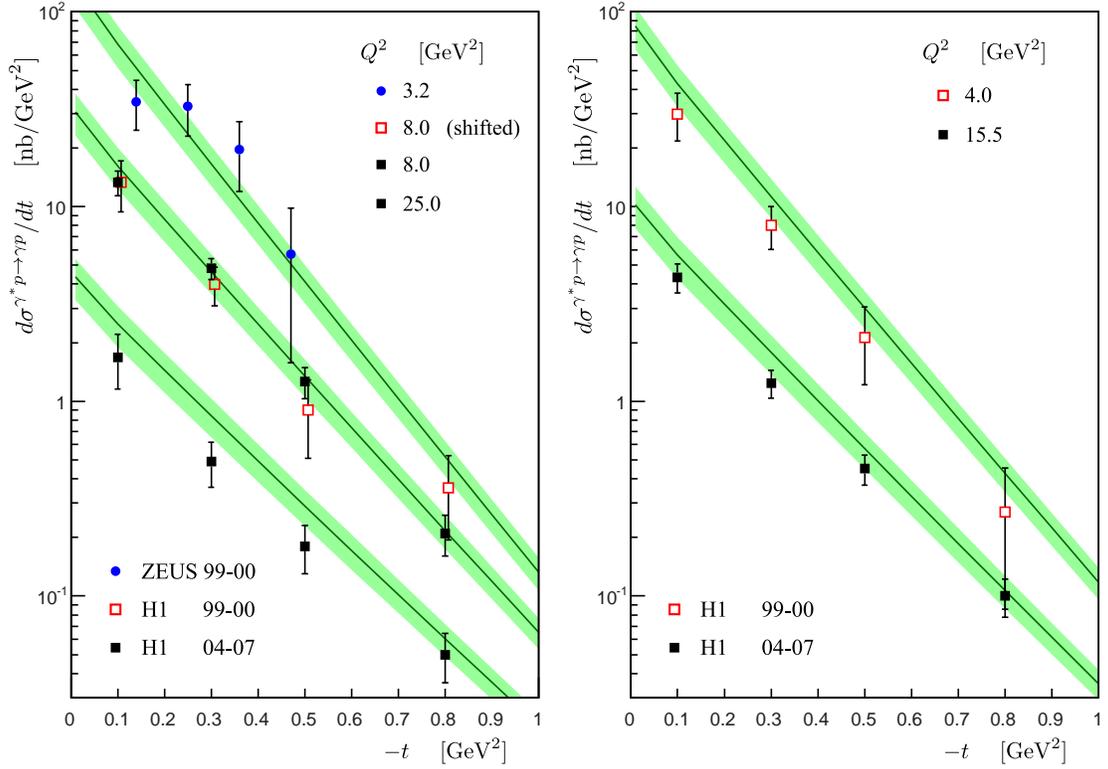}
\end{center}
\caption{Differential DVCS cross section versus $−t$ for a set of $Q^2$
  values 
  and large $W$ values ranging from 71~GeV at low $Q^2$ to 104~GeV at the highest $Q^2$.
  Data are taken from Refs.~\ci{Aktas:2005ty,h1-dvcs,Chekanov:2003ya,zeus-dvcs}, where 
  statistical and systematical errors are added in quadrature and normalization 
  uncertainties were ignored.  
  Our predictions are shown as solid lines with errors represented by shadowed bands.}
\label{fig:hera-dvcs}
\end{figure}

The H1 \cite{Aktas:2005ty,h1-dvcs} and ZEUS \cite{Chekanov:2003ya,zeus-dvcs}  
collaborations published $t$-differential $\gamma^* p \rightarrow \gamma p$
cross section. The kinematics for these data is characterized by small
$\xbj$, typically $10^{-3}- 10^{-4}$, a photon virtuality that varies between
3 and $25\,\gev^2$ and large $W$, of order $100\,\gev$. In this kinematical
range the GPDs at small skewness and small $x$ control the DVCS cross section. 
The dominant contribution to the cross sections comes from the GPD $H$.
The GPD $E$ is suppressed, see Tab.\ \ref{table:CFFdep}. The contribution from 
$\widetilde H$, although not suppressed as compared to $H$ (see Tab.\ \ref{table:CFFdep}),
is negligible here since $\widetilde{H}$ is much smaller than $H$; this is 
in particular the case for the sea quark contribution. We remark that for the leptoproduction of vector mesons,
$\widetilde H$ does not contribute to leading-twist accuracy.
Due to the large range of $Q^2$ in which
the HERA data are available, evolution of the GPD $H$ plays a decisive role.
Using the scale-dependent parametrization of $H$ introduced in Sect.\ 
\ref{subsec:2.1.1} we evaluate the DVCS cross section at HERA kinematics to 
leading-twist accuracy and LO of perturbative QCD. As the factorization scale
we choose $\mu_F=Q$. Our results are compared to the HERA data in Fig.\
\ref{fig:hera-dvcs}. Reasonable agreement with experiment is achieved within
experimental errors and theoretical uncertainties. A similar observation has
also been made in \ci{mes11}. In this paper only the GPD $H$ is taken into account
and parametrized in terms of an SO(3) $t$-channel partial wave expansion. 
A combined fit of the GPD parameters to the HERA DVCS and DVMP data is performed. 
In contrast to the analysis carried through
in \ci{GK1,GK3} (see Sect.\ \ref{subsec:2.2}) DVMP is also computed within the
collinear factorization approach. This necessitates a GPD that differs from the one
we are using in particular at the cross-over line $\xi=x$, and that leads to very strong evolution effects. The quality 
of the combined fit to the HERA data on DVCS and DVMP performed in \ci{mes11} is 
comparable to that of our results.

\section{Future Experiments}
\label{sec:4}

\begin{figure}[tp]
\begin{center}
\includegraphics[width=0.9\tw]{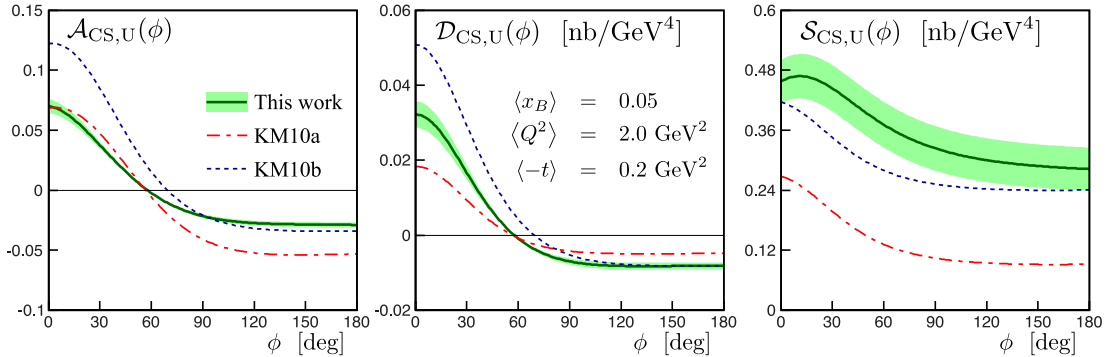}
\end{center}
\caption{Charge-spin asymmetry (left), cross section difference (middle) and
  sum (right) as a function of $\phi$ for one of the COMPASS-II kinematical
  bins at $x_B=0.05$, $Q^2=2$~GeV$^2$ and $-t=0.2$~GeV$^2$. Our predictions are shown
as solid lines with error bands, the predictions for the two scenarios given in
Ref.~\ci{kume09} are represented by short dashed and dash-dotted lines.}
\label{fig:Compass}
\end{figure}

The COMPASS collaboration will have a two-week DVCS run in 2012 with a 160~GeV
muon beam, followed by a longer run in 2015 and later
\ci{COMPASSproposal}. They plan to measure the $t$-slope of the
$\phi$-integrated photon electroproduction cross section, as well as 
specific charge and spin observables. The polarized muon beam is produced
through pion decay, its polarization therefore changes sign when the beam
charge is reversed, {\it
  i.e.} $\mu^+$ and $\mu^-$ are polarized along opposite directions. COMPASS plans
to measure mixed charge-spin (CS) cross sections differences and sums defined as
follows:
\ba
\mathcal S_{\rm CS,U} ( \phi ) & \equiv & d\sigma^{\stackrel{+}{\rightarrow}} + d\sigma^{\stackrel{-}{\leftarrow}} 
\=  2 d\sigma_{UU} ( 1 - A_{\rm LU, I}( \phi ) ) \nn\\
\mathcal D_{\rm CS,U} ( \phi ) & \equiv & d\sigma^{\stackrel{+}{\rightarrow}} - d\sigma^{\stackrel{-}{\leftarrow}}
\=  2 d\sigma_{UU} ( A_{\rm C}( \phi ) - A_{\rm LU, DVCS}( \phi )  ) \nn\\
\mathcal A_{\rm CS,U} ( \phi ) & \equiv &  \frac{\mathcal D_{\rm CS,U}}{\mathcal S_{\rm CS,U}} 
\=  \frac{ A_{\rm C}( \phi ) - A_{\rm LU, DVCS}( \phi ) }{1 - A_{\rm LU, I}( \phi )} 
\ea

\noindent where the cross sections $d\sigma^{h_\mu,e_\mu}$ are defined in Eq.\ 
\req{eq-airapetian-asymmetries}.
Predictions from our approach for these three observables are shown in 
Fig.~\ref{fig:Compass} along with predictions made in Ref.\ \ci{kume09}.

\begin{figure}[bp]
\begin{center}
\includegraphics[width=0.8\tw]{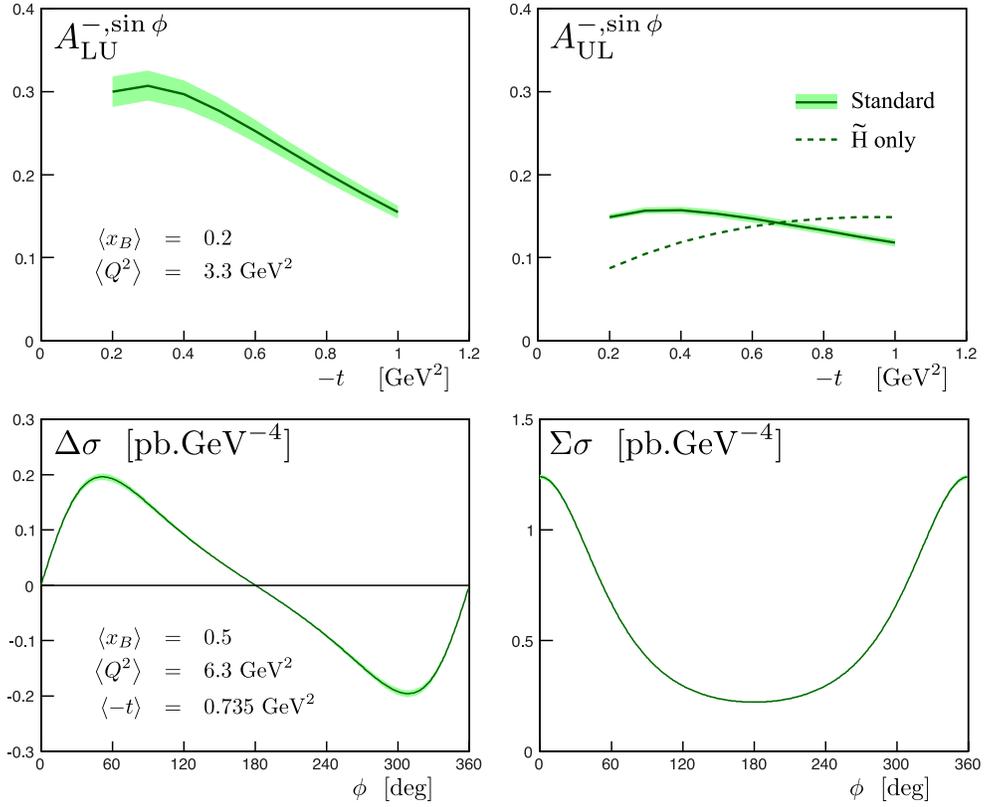}
\end{center}
\caption{Upper left and right: the $\sin\phi$ harmonics of the beam and target
  spin asymmetries $A_{\rm LU}^{\sin\phi}$ and $A_{\rm UL}^{\sin\phi}$  versus
  $-t$ for a typical bin accessible with CLAS12
  \ci{CLAS12prop}. Lower left and right: the difference and sum
  of the helicity  cross sections versus $\phi$ in one of the
  kinematical bin of the Hall~A DVCS experiment
  \ci{HallA12prop}. Our predictions are shown as solid lines with 
  error bands.}
\label{fig:JLab12}
\end{figure}

The CLAS12 collaboration plans to measure the beam and
target spin asymmetries as well as cross sections for DVCS with the 11~GeV
electron beam as soon as 2015~\ci{CLAS12prop}. In 
Fig.~\ref{fig:JLab12} our predictions for the $\sin\phi$ moments of the
beam and target spin asymmetries $A_{\rm LU}^{-,\sin\phi}$ and 
$A_{\rm  UL}^{-,\sin\phi}$, evaluated with the help of the analogue of 
\req{eq:projection}, are displayed for a typical kinematical bin accessible
with CLAS12. The Hall~A collaboration will also run a DVCS experiment very 
soon after the 12~GeV upgrade is complete~\ci{HallA12prop}. Our predictions
for  the difference and sum of the helicity cross sections are shown in 
Fig.~\ref{fig:JLab12} for one of the planned kinematical settings. 

More detailed predictions can be obtained from the authors on request.

\section{Summary and outlook}

Since 2000 there has been a great wealth of measurements related to the DVMP
and DVCS processes. Factorization theorems assert that these hard exclusive
processes can be interpreted in terms of partonic degrees of freedom, provided 
the virtuality of the
exchanged photon is large enough. In that case the partonic picture involves
GPDs, which are universal, process-independent quantities. Up to now, 
both processes have mostly been studied independently of each other, and it 
is therefore an important check of consistency to test whether the GPDs used
to describe one channel can be used to describe the other channel as well. 
Obviously universality is exploited at the most in a combined analysis of
both DVMP and DVCS but this large-scale program is beyond the scope of the
present article.

A necessary and interesting first step in exploiting universality is the use of
a set of GPDs extracted from an analysis of DVMP data from H1, ZEUS, E665, COMPASS
and HERMES \ci{GK3,GK5,GK4} in a then parameter-free evaluation of DVCS and the
detailed comparison with the available $ep\to ep\gamma$ data. The
various observables are computed within an approach in which the DVCS
amplitudes are calculated to leading-twist accuracy and leading-order
perturbative QCD while, similarly to Ref.~\cite{Belitsky:2010jw},
the leptonic tensor is taken into account without any 
approximation. We stress that DVCS is treated in the collinear factorization framework. This is
consistent with the calculation of DVMP as described in Sect.~\ref{subsec:2.2} and hence, the GPDs
extracted from DVMP can be used for the calculation of DVCS observables. In both processes
the quarks are emitted and reabsorbed from the proton collinearly with the proton momenta.

We observe very good overall agreement between our predictions and most of 
the H1, ZEUS and HERMES data but a less satisfactory description of the large $\xbj$ (small
$W$) data from Jefferson Lab, where power corrections might be large.
We notice that the most recent HERMES measurements~\ci{Airapetian:2012pg}
using a recoil detector in order to achieve fully 
exclusive final states, give a significantly larger beam spin asymmetry 
(in absolute value). The recoil data are in perfect agreement with our
results. Such dilution effect may be present in other HERMES observables as
well and estimates of this effect would be highly valuable.
The COMPASS, CLAS12 and Hall~A collaborations have planned to take DVCS data
in the coming years. In view of the good agreement with the available data 
demonstrated in our study we also give predictions for their main observables.

Our study makes it clear that the set of GPDs we are using which is 
extracted from DVMP, describes DVCS data over a large kinematical range rather
well in general although not perfectly in all details. This  is partly a
consequence of the different kinematic ranges of DVCS measurements and of the DVMP 
data used to constrain our GPD model. In addition, the GPD parametrization is of 
course an approximation. Thus, improvements of the GPD parametrization are 
required on the long run. In order to do so future data from COMPASS and JLab12 
will be of help. Possible improvements may include the use of  more recent 
versions of the PDFs, the proper scale dependence of the GPDs, an updated form factor analysis \ci{DK12}, eventual 
modifications of the profile functions of the double distribution parameterization 
and allowance for a non-zero $D$-term.

\section*{Acknowledgments}
The authors would like to thank M.~Diehl, N.~D'Hose, D.~M\"uller, W.-D.~Nowak
and G.~Schnell for many fruitful discussions and valuable inputs. 

This work was supported in part by the Commissariat \`a l'Energie Atomique et
aux Energies Alternatives and the GDR 3034 PH-QCD and by the BMBF under 
contract 06RY258.

\end{document}